\documentclass[12pt,a4paper]{article}

\usepackage[utf8]{inputenc}
\usepackage[T1]{fontenc}
\usepackage{amsmath,amssymb,amsfonts}
\usepackage{graphicx}
\usepackage{booktabs}
\usepackage{multirow}
\usepackage{array}
\usepackage{caption}
\usepackage{subcaption}
\usepackage[round,authoryear]{natbib}
\usepackage{hyperref}
\usepackage{geometry}
\usepackage{setspace}
\usepackage{threeparttable}
\usepackage{float}
\usepackage{placeins}
\graphicspath{{figures/}}

\newcommand{\RV}{\mathrm{RV}}
\newcommand{\RVPK}{\mathrm{RV}^{PK}}
\newcommand{\VIX}{\mathrm{VIX}}

\newcommand{\E}{\mathbb{E}}
\newcommand{\dhat}{\hat d}

\title{\textbf{Memory, Roughness, and Information Persistence in Financial
Markets:\\[3pt] A Structural Approach to Volatility Forecasting}}

\author{
Akash Deep\thanks{Department of Mathematics and Statistics, Texas Tech
University. Email: akash.deep@ttu.edu.} \and
Nicholas Appiah\thanks{Department of Mathematics and Statistics, Texas
Tech University.} \and
Svetlozar T. Rachev\thanks{Department of Mathematics and Statistics, Texas
Tech University.}
}

\date{April 2026}

\begin{document}

\maketitle

\begin{abstract}
\noindent This paper studies the joint role of long-memory dynamics,
rough-volatility behavior, and persistence-based forecasting features in
equity volatility modeling. We combine semiparametric long-memory
estimation, rough-volatility diagnostics, and structured forecasting
regressions to examine whether persistence measures contain economically
meaningful forecasting information beyond conventional volatility
predictors. Using a panel of 115 S\&P~500 constituents from November 2001
through April 2026, we document that volatility proxies exhibit substantial
long-memory behavior and locally rough dynamics. The cross-sectional mean
Geweke--Porter-Hudak estimate of the memory parameter is $\dhat = 0.226$,
while the corresponding local-Whittle estimate is $\dhat = 0.440$, with
statistical significance observed across nearly the entire panel. Rolling
estimates of persistence rise substantially during the global financial
crisis and the COVID period and display a positive contemporaneous
association with the VIX.
We then examine whether persistence-related features improve out-of-sample
volatility forecasts beyond standard HAR and HAR-X benchmarks.
Incorporating cross-sectional persistence aggregates, sectoral persistence
measures, and persistence-by-stress interaction terms produces moderate but
statistically significant forecasting improvements, particularly at longer
horizons and during stress regimes. Forecast gains are strongest during
periods of elevated market volatility and in volatility-managed portfolio
applications. The results suggest that persistence measures may serve as
useful reduced-form indicators of the duration and propagation of
uncertainty in financial markets, although the paper does not claim
structural identification of the economic mechanisms generating persistence.
\end{abstract}



\newpage

\section{Introduction}
\label{sec:intro}

Volatility forecasting occupies a central role in financial econometrics
because volatility enters risk management, derivative pricing, portfolio
allocation, margin determination, stress testing, and macro-financial
monitoring. A large empirical literature has documented that volatility
measures exhibit substantially greater persistence than raw asset returns.
Classical long-memory models formalized this observation by extending
conventional weak-dependence frameworks to allow autocorrelations that
decay at hyperbolic rather than exponential rates
\citep{granger1980introduction,hosking1981fractional}. Subsequent work
incorporated fractional dependence into conditional-volatility specifications
through FIGARCH-type models \citep{baillie1996fractionally}, while the
realized-volatility literature improved the empirical measurement of latent
volatility dynamics using high-frequency data
\citep{andersen2003modeling,barndorff2002econometric}.

Despite extensive empirical evidence for persistence in volatility proxies,
the interpretation of persistence remains controversial. One branch of the
literature views long-memory behavior as reflecting structural
characteristics of financial markets, including heterogeneous trading
horizons, slow information diffusion, and institutional frictions
\citep{muller1997volatilities}. Another branch argues that apparent
long-range dependence may arise from structural breaks, stochastic regime
changes, or nonstationary volatility dynamics rather than genuine fractional
dependence \citep{mikosch2004nonstationarities}. As a result, the economic
interpretation of persistence measures remains unsettled. The central issue
is therefore not simply whether a fractional parameter can be estimated, but
whether persistence measures behave coherently across market states and
contain useful information for forecasting and risk management.

This paper examines volatility persistence from a reduced-form perspective.
Rather than interpreting the memory parameter as a purely technical object
associated with fractional differencing, we investigate whether rolling
persistence measures behave empirically like indicators of the temporal
propagation of uncertainty in financial markets. More specifically, we study
whether persistence measures rise during sustained stress episodes, whether
they co-move with broad market uncertainty indicators, and whether they
improve volatility forecasts beyond standard benchmark models. The analysis
does not attempt to identify a unique structural mechanism generating
persistence. Instead, persistence is treated as an empirical summary
statistic whose dynamics may reflect several overlapping mechanisms,
including heterogeneous investor horizons, liquidity conditions, delayed
information processing, regime shifts, and market-wide stress propagation.

The paper also considers the relationship between long-memory behavior and
rough volatility. Recent research has shown that volatility paths often
display substantial local irregularity, with estimated Hurst exponents
significantly below the Brownian benchmark of $\tfrac12$
\citep{gatheral2018volatility,bennedsen2021decoupling}. Although roughness
and long memory are sometimes presented as competing explanations of
volatility dynamics, they describe different temporal characteristics. Long
memory concerns the persistence of dependence over long horizons, whereas
roughness concerns local path irregularity at very short horizons. The
empirical analysis in this paper suggests that both characteristics may
coexist within the same volatility process.

A further issue concerns the role of machine learning in volatility
forecasting. Flexible forecasting algorithms can capture nonlinear
interactions and high-dimensional predictor structures that may be difficult
to represent using conventional parametric models. However, purely
data-driven forecasting systems often lack economic interpretability and may
produce unstable results outside the estimation sample
\citep{gu2020empirical}. The present paper adopts an intermediate approach:
econometric methods are used to construct interpretable persistence-related
features, while forecasting systems incorporate these features alongside
conventional volatility predictors and market-state variables.

The empirical analysis uses a panel of 115 S\&P~500 stocks over the period
November 2001 through April 2026. We estimate rolling long-memory measures
using both the Geweke--Porter-Hudak and local-Whittle estimators and
estimate local roughness using rolling Hurst exponents. These variables are
incorporated into a layered forecasting framework that compares HAR, HAR-X,
persistence-augmented regressions, shrinkage estimators, and tree-based
machine-learning methods.

The empirical results may be summarized as follows. First, volatility
proxies exhibit substantial long-memory behavior and strong local roughness
throughout the panel. Second, rolling persistence measures rise markedly
during periods of market stress and display positive association with the
VIX. Third, persistence-related forecasting features provide statistically
significant incremental improvements beyond HAR and HAR-X benchmarks,
particularly during stress regimes and at longer horizons. Fourth, the gains
arise primarily from cross-sectional and interaction effects rather than
from own-stock persistence alone. Finally, linear persistence-augmented
systems outperform tree-based machine-learning methods, suggesting that
economically structured feature design matters more than estimator
complexity.

The remainder of the paper proceeds as follows. Section~\ref{sec:literature}
reviews the literature. Section~\ref{sec:theory} discusses the theoretical
foundations of persistence. Section~\ref{sec:econometrics} presents the
econometric framework. Section~\ref{sec:structural} develops the
reduced-form interpretation. Section~\ref{sec:ml} introduces the forecasting
framework. Section~\ref{sec:data} describes the data and empirical design.
Section~\ref{sec:results} presents the results. Section~\ref{sec:economic}
discusses the economic implications. Section~\ref{sec:conclusion} concludes.

\section{Literature Review: Persistence, Roughness, and Volatility}
\label{sec:literature}

The literature on volatility persistence originates with foundational work
on fractional integration and ARFIMA processes, which extended classical
time-series analysis by allowing autocorrelation structures with
hyperbolically decaying dependence rather than conventional exponential
decay \citep{granger1980introduction,hosking1981fractional}. These
developments became especially important in financial econometrics because
volatility proxies such as squared returns, absolute returns, and realized
variance measures exhibit substantially stronger persistence than raw asset
returns themselves \citep{ding1993long}. The resulting evidence motivated
the development of models capable of capturing long-range dependence in
financial volatility while remaining consistent with broader conditional
heteroskedasticity frameworks.

One important contribution in this direction was the FIGARCH model
introduced by \citet{baillie1996fractionally}, which incorporated
fractional integration directly into conditional variance dynamics.
FIGARCH generalized conventional GARCH specifications by allowing shocks to
decay at hyperbolic rates, thereby accommodating the empirically observed
persistence of volatility. The model rapidly became a standard benchmark in
long-memory volatility research because it provided a tractable parametric
representation of persistent conditional heteroskedasticity. Subsequent
extensions introduced asymmetric responses, multivariate formulations, and
alternative fractional structures, further expanding the long-memory
volatility literature.

At approximately the same time, the realized-volatility literature
substantially improved the empirical measurement of latent volatility
processes. \citet{andersen2003modeling} and \citet{barndorff2002econometric}
demonstrated that high-frequency financial data could be used to construct
more accurate ex post measures of integrated volatility. These developments
strengthened empirical evidence for volatility persistence because
realized-volatility measures contain less measurement noise than traditional
squared-return proxies. The resulting data environment allowed researchers
to study persistence properties with considerably greater precision.

Another important development was the HAR model proposed by
\citet{corsi2009simple}. Rather than imposing exact fractional integration,
the HAR framework approximates long-memory behavior using a parsimonious
multi-horizon structure involving daily, weekly, and monthly
realized-volatility components. The HAR model proved empirically successful
because it combines forecasting accuracy with economic interpretability. In
particular, it is consistent with the idea that financial markets contain
heterogeneous agents operating at different investment horizons
\citep{muller1997volatilities}. The HAR framework remains one of the most
widely used benchmarks in volatility forecasting.

Despite substantial empirical support for persistence, the interpretation of
persistence remains controversial. One branch of the literature views long
memory as reflecting genuine structural characteristics of financial markets,
including heterogeneous trading frequencies, delayed information diffusion,
and institutional frictions. Another branch argues that apparent long-range
dependence may arise from structural breaks, regime changes, stochastic
volatility shifts, or nonlinear dynamics rather than from genuine fractional
integration \citep{mikosch2004nonstationarities}. This critique implies that
estimated persistence parameters should be interpreted cautiously because
they may reflect evolving market conditions rather than stable stochastic
mechanisms.

The debate surrounding long memory became even more important following the
emergence of the rough-volatility literature. \citet{gatheral2018volatility}
showed that volatility paths frequently exhibit substantial local
irregularity, with estimated Hurst exponents significantly below the
Brownian-motion benchmark of $H=\tfrac12$. Subsequent work clarified that
roughness and long memory are not necessarily competing concepts.
\citet{bennedsen2021decoupling} argued that short-run roughness and
long-horizon persistence may represent distinct temporal features of
volatility processes. The empirical analysis in this paper adopts precisely
this perspective by estimating both rolling persistence measures and rolling
Hurst exponents within the same forecasting framework.

In parallel with these econometric developments, machine learning has become
increasingly important in volatility forecasting and empirical asset pricing.
Flexible learning algorithms are capable of capturing nonlinear interactions,
threshold effects, and high-dimensional predictor structures
\citep{breiman2001random,friedman2001greedy,gu2020empirical,hastie2009elements}.
However, purely data-driven approaches may produce unstable results outside
the estimation sample and frequently lack economic interpretability.

This tension between statistical flexibility and economic interpretation has
become a central issue in modern empirical finance. Recent research
increasingly emphasizes hybrid approaches in which economically motivated
features are combined with flexible forecasting architectures. In this
setting, the quality and interpretability of the feature space may be more
important than the complexity of the forecasting algorithm itself.

Taken together, the literature reveals several unresolved issues. First,
there remains substantial disagreement regarding the economic interpretation
of volatility persistence and its relationship to regime dependence,
structural breaks, and heterogeneous trading behavior. Second, the
relationship between long-memory dynamics and rough-volatility behavior is
still not fully understood. Third, relatively few studies examine whether
persistence measures themselves contain incremental forecasting information
beyond conventional volatility predictors and market-state variables. The
present paper contributes by examining persistence measures from a
reduced-form empirical perspective, studying whether rolling persistence
measures behave coherently across market regimes and provide economically
meaningful forecasting information.

\section{Theoretical Foundations of Long Memory and Information
Persistence}
\label{sec:theory}

The concept of long memory emerged from attempts to describe stochastic
processes whose dependence structures cannot be adequately represented by
conventional short-memory models. Long-memory processes retain measurable
dependence over substantially longer horizons, typically characterized
through a slowly decaying autocovariance structure in which dependence
declines at a hyperbolic rather than exponential rate
\citep{granger1980introduction,hosking1981fractional}.

Formally, a covariance-stationary long-memory process may be characterized
by an autocovariance function satisfying
\begin{equation}
  \gamma(k) \sim c_\gamma\, k^{2d-1}\quad \text{as } k\to\infty,
  \label{eq:lm-autocov}
\end{equation}
where $d\in(0,\tfrac12)$ denotes the memory parameter. Because $2d-1>-1$,
the autocovariances are not absolutely summable, implying persistent
dependence across long horizons. Equivalent characterizations arise in the
frequency domain,
\begin{equation}
  f(\lambda) \sim c_f\, \lambda^{-2d}, \quad \lambda \to 0,
\end{equation}
providing the foundation for semiparametric estimators of long memory,
including the Geweke--Porter-Hudak and local-Whittle procedures employed
later in the paper \citep{robinson1995gaussian}.

The fractional-difference operator $(1-L)^d$ provides the bridge between
these probabilistic properties and empirically estimable time-series models.
Fractionally integrated ARFIMA systems express current observations as
weighted combinations of past innovations, where the weights decay gradually
rather than disappearing rapidly. Relative to standard ARMA structures,
fractional integration allows shocks to persist over much longer horizons
while preserving covariance stationarity when $d<\tfrac12$. In financial
econometrics, this framework proved especially useful because volatility
measures routinely display persistence patterns that are inconsistent with
short-memory dynamics.

Although long-memory models originated as statistical descriptions of
temporal dependence, their interpretation in financial applications
naturally raises broader economic questions. When persistence is observed in
volatility proxies rather than in returns themselves, the memory parameter
may be viewed as summarizing the duration over which uncertainty shocks
continue to influence market behavior. Importantly, this interpretation
should not be understood as structural identification of a unique mechanism.
Rather, persistence measures are best viewed as reduced-form summaries
reflecting several overlapping economic forces that jointly influence the
temporal propagation of volatility.

One important explanation for persistent volatility dynamics arises from
heterogeneous market participation. Financial markets contain agents
operating at different horizons, using different information sets, and
responding at different speeds to new information. High-frequency traders
may react within milliseconds to order-flow imbalances and liquidity
conditions, whereas institutional investors may rebalance portfolios over
days, weeks, or months. Pension funds, insurance companies, and
macro-oriented asset managers frequently adjust positions gradually in
response to evolving risk conditions rather than immediately after news
arrivals. The aggregation of these heterogeneous responses can generate
long-horizon persistence even when individual agents themselves do not
follow fractional stochastic laws
\citep{muller1997volatilities,corsi2009simple}.

Another important channel involves delayed information processing and
gradual uncertainty resolution. Macroeconomic announcements,
monetary-policy surprises, geopolitical shocks, or systemic financial
disturbances may become publicly observable immediately, yet their
implications for risk, liquidity, hedging demand, and portfolio reallocation
may unfold only gradually. Under such conditions, volatility persistence
reflects not only the arrival of information but also the prolonged process
through which financial markets absorb, interpret, and respond to
uncertainty. In this setting, persistence measures may summarize the speed
at which uncertainty dissipates throughout the financial system.

Market microstructure mechanisms provide an additional source of
persistence. Order splitting, inventory management by dealers, intermittent
liquidity provision, and interactions between informed and uninformed
traders can generate volatility clustering at short horizons. Repeated
microstructure frictions may then accumulate into broader dependence
patterns observable over longer periods. Under this interpretation,
long-memory behavior represents the macro-level consequence of repeated
local frictions and trading adjustments rather than a purely abstract
time-series property.

Behavioral mechanisms may also contribute to persistent volatility
dynamics. Financial markets do not always process information
instantaneously or efficiently. Underreaction, overreaction, limited
attention, ambiguity aversion, and delayed belief updating can all produce
gradual adjustment processes in which shocks remain economically relevant
for extended periods. Although the classical long-memory literature was not
originally formulated within behavioral-finance frameworks, the empirical
phenomenon of slow volatility adjustment is consistent with several
behavioral explanations of market dynamics.

These considerations suggest that persistence measures may serve as
reduced-form indicators of the temporal organization of uncertainty in
financial markets. A stylized representation may be written as
\begin{equation}
  I_t \;=\; \sum_{k=0}^{\infty} w_k\, \varepsilon_{t-k}, \qquad
  w_k \sim c\, k^{d-1},
  \label{eq:info-persistence}
\end{equation}
where $I_t$ denotes an uncertainty-related process influenced by past
informational shocks $\varepsilon_{t-k}$, and $d$ governs the speed at
which the influence of past shocks declines over time.

It is important, however, to distinguish carefully between statistical
persistence and economic interpretation. The presence of estimated long
memory does not necessarily imply that markets literally follow stationary
fractional stochastic laws. Structural breaks, regime changes, stochastic
volatility shifts, and evolving institutional environments may generate
empirical behavior resembling long-range dependence. Consequently,
persistence measures should not be interpreted mechanically as identifying
a single structural mechanism. Instead, they should be viewed as empirical
indicators whose behavior may reflect several interacting economic channels
simultaneously.

The interpretation developed in this paper therefore remains intentionally
reduced-form. The central empirical question is not whether a unique
structural mechanism can be identified, but rather whether persistence
measures behave coherently across market states and whether they contain
useful forecasting information beyond conventional volatility predictors.
If persistence measures rise systematically during sustained stress
episodes, co-move with broad uncertainty indicators, and improve volatility
forecasts during crisis periods, then they may provide economically
informative summaries of market conditions even without full structural
identification.

This perspective also accommodates the rough-volatility literature.
Roughness and long memory describe different aspects of temporal dependence
rather than mutually exclusive mechanisms. Long memory concerns
low-frequency dependence and the persistence of shocks across longer
horizons. Roughness concerns local path irregularity at very short horizons.
A rough-volatility process satisfies
\begin{equation}
  \E\bigl|\log\sigma^2_{t+\Delta} - \log\sigma^2_t\bigr|^q
  \;\propto\; \Delta^{qH},
  \label{eq:rough-scaling-sec3}
\end{equation}
where $H<\tfrac12$ implies rougher behavior than standard Brownian motion
\citep{gatheral2018volatility}. A volatility process may therefore be
simultaneously rough at short horizons and persistent at longer horizons.
Financial markets may react sharply and irregularly immediately after news
arrivals while still exhibiting prolonged uncertainty propagation over
subsequent periods.

The coexistence of roughness and persistence suggests that volatility
dynamics should be analyzed across multiple temporal scales rather than
through a single dependence measure. High-frequency volatility fluctuations
may primarily reflect microstructure effects and rapid information arrivals,
whereas medium- and long-horizon persistence may reflect broader
institutional adjustment processes, funding conditions, and macro-financial
uncertainty propagation. The empirical framework developed later in the
paper incorporates both dimensions simultaneously through rolling estimates
of long-memory parameters and rolling Hurst exponents.

Finally, the interpretation of persistence developed here has practical
implications for volatility forecasting and risk management. If persistence
measures summarize the expected duration of uncertainty conditions, then
they may contain forecasting information that differs conceptually from
conventional volatility-level indicators such as the VIX. Two periods with
similar contemporaneous volatility levels may nevertheless differ
substantially in the expected duration of uncertainty propagation.
Persistence measures may therefore provide complementary information
regarding how long elevated volatility conditions are likely to persist.

The empirical sections of the paper investigate this possibility directly
by incorporating persistence-related features into structured forecasting
systems and examining whether they improve forecasting performance across
market regimes, sectors, and forecast horizons.

\section{Econometric Modelling of Fractional and Rough Volatility}
\label{sec:econometrics}

The econometric analysis of volatility persistence has evolved along two
closely related but conceptually distinct directions. The first focuses on
long-memory dynamics and emphasizes slow hyperbolic decay over medium and
long horizons. The second focuses on rough volatility and emphasizes highly
irregular local behavior at very short horizons. The long-memory literature
developed primarily within discrete-time econometrics through fractional
integration and ARFIMA-type models. Rough-volatility models, by contrast,
emerged from continuous-time stochastic-volatility theory and high-frequency
financial econometrics. Recent empirical evidence suggests that these two
perspectives are not mutually exclusive. The empirical framework developed
in this paper therefore incorporates both dimensions in a unified forecasting
environment.

\subsection{ARFIMA, GARCH, and FIGARCH}

The ARFIMA framework provides a natural discrete-time representation of
fractional dependence: for a covariance-stationary process $\{X_t\}$,
\begin{equation}
  \phi(L)\, (1-L)^d\, X_t \;=\; \theta(L)\,\eta_t,
\end{equation}
where $\phi(L)$ and $\theta(L)$ are finite-order lag polynomials, $d$
denotes the fractional integration parameter, and $\eta_t$ is a white-noise
innovation process \citep{granger1980introduction,hosking1981fractional}.
The operator $(1-L)^d$ generalizes conventional integer differencing to
fractional orders and allows shocks to decay gradually over long horizons.
When applied to volatility proxies such as squared returns, absolute
returns, range-based volatility measures, or realized variance, ARFIMA
models provide parsimonious representations of long-range dependence in
financial volatility. Unlike standard short-memory models, fractional
integration allows dependence to persist over extended periods while
preserving covariance stationarity when $d<\tfrac12$.

The conditional heteroscedasticity literature approached volatility
persistence from a different perspective. Beginning with the ARCH model of
\citet{engle1982autoregressive} and the GARCH extension of
\citet{bollerslev1986generalized}, volatility dynamics were modeled directly
through evolving conditional variance equations. Standard GARCH systems
successfully capture volatility clustering but imply geometric decay in
the effects of shocks. Empirical volatility series, however, often display
substantially slower decay than conventional GARCH structures permit. The
FIGARCH model of \citet{baillie1996fractionally} addressed this limitation
by embedding fractional integration directly within the conditional-variance
recursion,
\begin{equation}
  \bigl[1 - \beta(L)\bigr]\, h_t \;=\; \omega
  + \bigl[1 - \beta(L) - \phi(L)(1-L)^d\bigr]\, \varepsilon_t^2 \,,
\end{equation}
where $h_t$ denotes conditional variance and $\varepsilon_t$ denotes return
innovations. FIGARCH allows volatility shocks to decay hyperbolically rather
than exponentially while preserving the conditional-heteroskedasticity
structure.

The FIGARCH framework became particularly influential because it provided an
explicit connection between long-memory econometrics and volatility
modeling. Numerous subsequent extensions introduced asymmetric effects,
multivariate specifications, nonlinear dynamics, and regime dependence.
Nevertheless, the central empirical motivation remained unchanged:
volatility shocks appear to dissipate much more slowly than conventional
short-memory volatility models predict.

An important practical issue concerns interpretation. Although ARFIMA and
FIGARCH systems estimate similar persistence behavior, they represent
different modeling philosophies. ARFIMA-type systems emphasize the
dependence structure of the observed process itself, whereas GARCH-type
systems emphasize the evolution of latent conditional variance. In practice,
both approaches frequently generate evidence consistent with persistent
volatility dynamics.

\subsection{Realized Volatility and HAR Models}

The development of realized-volatility methodology substantially improved
empirical analysis of volatility persistence. Traditional volatility proxies
based on squared daily returns contain substantial measurement noise and
therefore obscure underlying dependence structures. High-frequency financial
data made it possible to construct more accurate ex post measures of
integrated volatility through realized-variance estimators
\citep{andersen2003modeling,barndorff2002econometric}.
Realized-volatility measures significantly strengthened empirical evidence
for persistent volatility dynamics because they provide less noisy
approximations to latent volatility processes. These measures also
facilitated more precise estimation of long-memory behavior and improved
volatility forecasting performance across a wide range of applications.

An especially important contribution was the heterogeneous autoregressive
(HAR) model proposed by \citet{corsi2009simple}. The HAR framework
approximates long-memory behavior using a parsimonious multi-horizon
structure rather than imposing exact fractional integration,
\begin{equation}
  \log \mathrm{RV}_{t+1} \;=\; \beta_0 + \beta_d \log \mathrm{RV}^{(d)}_t
  + \beta_w \log \mathrm{RV}^{(w)}_t
  + \beta_m \log \mathrm{RV}^{(m)}_t
  + u_{t+1},
\end{equation}
where $\mathrm{RV}^{(d)}_t$, $\mathrm{RV}^{(w)}_t$, and
$\mathrm{RV}^{(m)}_t$ denote daily, weekly, and monthly realized-volatility
components. The HAR model proved remarkably successful empirically because
it captures many observed features of long-memory volatility dynamics while
remaining computationally simple and economically interpretable. The
framework is closely connected to heterogeneous-agent theories in which
market participants operate at different investment horizons
\citep{muller1997volatilities}. Daily components capture short-horizon
trading activity, whereas weekly and monthly components reflect
slower-moving institutional and macroeconomic adjustments.

The HAR framework also became a dominant benchmark in volatility forecasting
because of its strong empirical performance relative to more complex
nonlinear specifications. In the present paper, HAR models serve as the
baseline forecasting architecture against which persistence-augmented models
are evaluated.

\subsection{Rough Volatility}

The rough-volatility literature introduced a complementary perspective on
volatility dynamics by emphasizing local path irregularity rather than
long-horizon persistence. \citet{gatheral2018volatility} showed that
volatility paths estimated from high-frequency financial data often display
substantially rougher behavior than implied by standard Brownian-motion
models. In particular, estimated Hurst exponents frequently lie well below
the classical benchmark $H=\tfrac12$.

A simplified rough-volatility representation may be written as
\begin{equation}
  \log \sigma_t \;=\; \mu + \nu\, B^H_t,
\end{equation}
where $B^H_t$ denotes fractional Brownian motion with Hurst exponent
$H<\tfrac12$. From an econometric standpoint, roughness is commonly
characterized through scaling relations involving moments of volatility
increments,
\begin{equation}
  \E|\log\sigma^2_{t+\Delta} - \log\sigma^2_t|^q \;\propto\; \Delta^{qH},
\end{equation}
where $H$ governs the scaling behavior of local volatility fluctuations
across short horizons. We estimate $H_t$ by the slope of
$\log m(q,\Delta)$ on $\log\Delta$ over a small grid of lags, divided
by $q$.

The rough-volatility framework proved highly successful in explaining
several empirical features of financial markets, including short-horizon
volatility behavior and implied-volatility surface dynamics
\citep{bennedsen2021decoupling}. Rough models also provided improved
consistency with observed option-market data and generated substantial
interest in both academic and practitioner communities.

Importantly, roughness and long memory characterize different temporal
properties of volatility. Roughness concerns highly local short-term
irregularity, whereas long memory concerns persistence across broader
temporal horizons. A volatility process may therefore be simultaneously
rough and persistent. The empirical framework adopted in this paper
explicitly incorporates both dimensions through rolling estimates of
long-memory parameters and rolling Hurst exponents.

\subsection{Semiparametric Estimation of Long Memory}

Empirical estimation of long memory presents substantial econometric
challenges because dependence structures must be distinguished from
nonstationarity, regime shifts, and structural breaks. The literature
therefore developed several semiparametric estimation procedures that focus
specifically on low-frequency behavior.

One widely used approach is the Geweke--Porter-Hudak (GPH)
log-periodogram estimator \citep{geweke1983estimation}. The estimator
exploits the low-frequency behavior of the spectral density by regressing
the logarithm of the periodogram on transformed Fourier frequencies. Using
the first $m$ low-frequency ordinates, the estimator takes the form
\begin{equation}
  \log I(\lambda_j) \;=\; \alpha
  - d\,\log\bigl[4\sin^2(\lambda_j/2)\bigr] + u_j,
\end{equation}
where $\dhat = -\hat\beta/2$ and bandwidth $m = \lfloor T^{0.65}\rfloor$.

Another important procedure is the local-Whittle estimator proposed by
\citet{robinson1995gaussian}. Rather than relying on regression, the
local-Whittle approach minimizes an approximation to the Gaussian likelihood
over the low-frequency band of the spectrum. The estimator generally
exhibits lower asymptotic variance under broad conditions and has become one
of the standard semiparametric procedures in long-memory econometrics. Both
estimators focus specifically on low-frequency dependence without imposing
complete parametric ARMA or GARCH structures, making them attractive for
empirical applications in which the short-memory component may be
misspecified or unstable.

In addition to semiparametric methods, the present paper also uses FIGARCH
estimates on representative subsets of stocks as parametric cross-checks.
Combining semiparametric and parametric approaches helps reduce dependence
on any single modeling framework and improves robustness of the empirical
conclusions.

\subsection{Identification Challenges}

Several identification problems complicate empirical interpretation of
persistence estimates. The first concerns structural nonstationarity.
Repeated regime shifts, evolving volatility states, and structural breaks
can generate empirical behavior resembling fractional dependence even when
the underlying process is not truly fractionally integrated
\citep{mikosch2004nonstationarities}. As a result, estimated persistence
parameters should not automatically be interpreted as evidence for
stationary long-memory laws.

The second challenge concerns the distinction between highly persistent
GARCH systems and genuine fractional integration. Near-integrated GARCH
models may produce empirical dynamics that are difficult to distinguish from
FIGARCH processes in finite samples, especially when persistence parameters
approach unity. A third challenge concerns rough-volatility estimation.
High-frequency financial data contain substantial market microstructure
noise arising from bid-ask bounce, asynchronous trading, discreteness
effects, and liquidity frictions. Unless realized-volatility measures are
carefully constructed, estimated roughness parameters may partly reflect
measurement distortions rather than genuine volatility dynamics.

The present paper addresses these issues through several robustness
procedures. Persistence estimates are computed using multiple volatility
proxies, including squared returns, Parkinson range-based measures, and
realized variance where available. Both GPH and local-Whittle estimators
are employed, together with FIGARCH cross-checks and rolling estimation
windows of varying lengths. The objective is not to establish exact
structural identification of fractional stochastic laws, but rather to
determine whether persistence-related measures behave coherently across
market states and contain useful forecasting information.

These identification issues are central to the broader interpretation
developed in the next section, where persistence measures are viewed as
reduced-form indicators of uncertainty propagation rather than as
mechanically interpreted fractional coefficients.

\section{Structural Interpretation of Memory Parameters}
\label{sec:structural}

The econometric literature traditionally introduces the memory parameter
$d$ as a statistical descriptor governing the decay rate of temporal
dependence. Within purely probabilistic frameworks, $d$ determines how
strongly past shocks continue to influence present observations. Once one
moves from abstract time-series theory to financial economics, however,
this interpretation becomes incomplete. Financial markets are institutional
systems in which information is generated, interpreted, delayed,
transmitted, amplified, and sometimes distorted by heterogeneous
participants. Persistence is not merely a property of a covariance function.
It is an empirical manifestation of how long the consequences of information
shocks remain economically active within the financial system.

This paper therefore interprets the rolling persistence estimate $\hat d_t$
as a reduced-form indicator of the expected duration of volatility shocks.
More specifically, $\hat d_t$ summarizes the speed at which financial
markets absorb informational disturbances and gradually return toward normal
uncertainty conditions. Persistence measures should not be interpreted as
identifying a single structural mechanism. Instead, they are treated as
empirical state variables reflecting the combined influence of several
interacting economic processes.

It is useful to distinguish carefully between statistical persistence and
economic persistence. Statistical persistence refers to the slow decay of
dependence in an observed series. Economic persistence refers to the
prolonged influence of shocks on market uncertainty, liquidity conditions,
risk perception, and portfolio adjustment. A macroeconomic announcement,
credit event, or geopolitical disturbance may occur at a single point in
time, yet its economic consequences may unfold gradually through changes in
expectations, portfolio rebalancing, collateral conditions, funding
constraints, and hedging demand. In this sense, volatility persistence
reflects not only the arrival of information but also the prolonged process
through which uncertainty is absorbed and resolved.

One important economic channel underlying persistence is information
diffusion. Financial markets process information across multiple temporal
horizons rather than through instantaneous adjustment. Even when prices
respond rapidly to public news, the implications of uncertainty may continue
to propagate through markets for extended periods. Market participants
differ substantially in sophistication, leverage, liquidity access, and
trading horizon. Consequently, adjustment to new information often occurs
sequentially rather than simultaneously.

A second channel involves heterogeneous market participation. Financial
systems contain agents operating at different frequencies and under
different institutional constraints. High-frequency traders respond to
order flow and liquidity conditions within milliseconds, whereas
institutional investors, pension funds, and asset managers may rebalance
positions over substantially longer horizons. The aggregation of
heterogeneous reactions across trading horizons can generate persistent
volatility dynamics even when individual participants themselves do not
follow fractional stochastic laws
\citep{muller1997volatilities,corsi2009simple}. Under this interpretation,
long memory represents the macroeconomic signature of micro-level
heterogeneity.

Liquidity conditions provide a third channel through which persistence may
arise. In highly liquid markets, shocks can often be absorbed relatively
quickly because market depth and risk-sharing capacity remain strong. In
stressed or illiquid markets, however, order imbalances may persist for
extended periods, spreads may widen, and volatility may remain elevated
long after the original informational disturbance occurs. Persistence
measures may therefore capture not only informational dynamics but also the
market's capacity to absorb risk.

Systemic stress and financial contagion constitute another important
mechanism. Disturbances originating in one market segment may propagate
gradually through collateral constraints, leverage adjustments, funding
markets, and cross-asset portfolio exposures. Under such conditions,
persistence reflects the duration of system-wide stress propagation rather
than isolated asset-specific volatility dynamics. This interpretation is
particularly relevant during major crisis periods such as the global
financial crisis and the COVID episode, when uncertainty remained elevated
across broad segments of the financial system for sustained periods.

Behavioral considerations also provide plausible explanations for
persistent volatility dynamics. Financial-market participants may exhibit
underreaction, delayed belief updating, ambiguity aversion, limited
attention, or gradual expectation adjustment. Under such conditions,
informational shocks are not fully incorporated into market behavior
immediately. Instead, volatility may remain elevated while beliefs and
positions adjust gradually over time.

These considerations motivate interpreting persistence measures as
state-dependent indicators of the temporal organization of uncertainty in
financial markets. A useful stylized representation may be written as
\begin{equation}
  I_t \;=\; \sum_{k=0}^{\infty} w_k\, \varepsilon_{t-k}, \qquad
  w_k \sim c\, k^{d-1},
\end{equation}
where $\varepsilon_{t-k}$ denotes past informational shocks. The parameter
$d$ governs the speed at which the influence of past disturbances declines
over time. Larger values of $d$ correspond to slower decay and therefore
longer persistence of uncertainty conditions.

Importantly, this interpretation remains intentionally reduced-form. The
paper does not claim that persistence estimates uniquely identify a single
economic mechanism. Persistence measures may simultaneously reflect
heterogeneous trading horizons, delayed information diffusion, liquidity
conditions, structural breaks, behavioral frictions, and systemic stress
propagation. Their usefulness lies not in exact structural identification
but in their ability to summarize economically meaningful temporal
characteristics of financial uncertainty.

This reduced-form interpretation is conceptually similar to the
interpretation of credit spreads in fixed-income markets. Credit spreads do
not isolate a single risk component but instead summarize several
overlapping influences, including default risk, liquidity premia,
macroeconomic uncertainty, and risk aversion. Likewise, persistence
measures summarize several interacting features of the financial information
environment rather than isolating one uniquely identifiable mechanism.

The empirical results developed later in the paper strongly support this
interpretation. The cross-sectional mean of rolling persistence estimates
rises sharply during major stress episodes, including the global financial
crisis and COVID. Moreover, persistence measures display substantial
positive association with the VIX and improve volatility forecasts
particularly during high-stress regimes and at longer forecast horizons.
These patterns are precisely what one would expect if persistence measures
capture the expected duration of unresolved uncertainty.

The interpretation developed here also helps clarify the relationship
between persistence and rough volatility. Roughness and long memory
characterize different temporal properties of volatility dynamics rather
than competing mechanisms. Roughness concerns local irregularity and
short-horizon path behavior, whereas persistence concerns the prolonged
influence of shocks across broader horizons. Financial markets may
therefore exhibit sharp local volatility fluctuations immediately after
news arrivals while simultaneously displaying prolonged uncertainty
propagation over longer periods.

This distinction is especially important for volatility forecasting.
Short-term volatility spikes and long-term persistence dynamics may contain
different types of predictive information. Conventional volatility-level
indicators such as the VIX primarily summarize the intensity of
contemporaneous uncertainty. Persistence measures, by contrast, may
summarize the expected duration of uncertainty conditions. Two periods with
similar volatility levels may nevertheless differ substantially in how long
elevated uncertainty is expected to persist.

From a forecasting perspective, persistence measures therefore provide
information distinct from traditional volatility proxies. A
high-persistence state implies not merely elevated volatility but also slow
expected resolution of uncertainty. This interpretation motivates the
forecasting framework developed in the next section, where
persistence-related variables are incorporated directly into structured
volatility forecasting systems.

\section{Structured Forecasting with Persistence-Based Features}
\label{sec:ml}

If persistence measures contain economically meaningful information, the
next question concerns how such information should be incorporated into
volatility forecasting systems. The approach adopted in this paper treats
persistence measures not as isolated econometric coefficients but as dynamic
state variables describing the temporal structure of financial uncertainty.
Forecasting systems are therefore constructed around economically
interpretable persistence-based features rather than around purely
mechanical lag structures.

The framework developed here occupies an intermediate position between
classical econometric modeling and purely data-driven machine learning.
Traditional parametric forecasting systems provide interpretability and
structural discipline but may fail to capture nonlinear interactions and
regime dependence. Purely flexible machine-learning systems may capture
complex patterns while sacrificing economic transparency and out-of-sample
stability \citep{gu2020empirical}. The objective of the present framework
is therefore not to replace econometric structure with machine learning,
but rather to combine economically interpretable feature construction with
flexible forecasting architectures.

The central idea is that persistence contains information distinct from
conventional volatility-level indicators. Traditional forecasting systems
such as HAR models primarily capture the intensity and temporal aggregation
of volatility. Market-state variables such as the VIX and MOVE indices
summarize contemporaneous stress conditions. Persistence measures
potentially contribute an additional dimension: the expected duration of
uncertainty propagation.

This distinction is economically important. Two periods may exhibit similar
contemporaneous volatility levels while differing substantially in how long
elevated uncertainty is expected to persist. One period may involve a
short-lived volatility spike associated with rapid uncertainty resolution,
whereas another may involve prolonged stress propagation across sectors,
funding markets, and institutional balance sheets. Persistence measures are
designed to capture this second dimension.

The forecasting framework therefore decomposes volatility dynamics into
three conceptually distinct channels. The first is the time-aggregation
channel captured by HAR-type dynamics. The second is the contemporaneous
stress-level channel captured by implied-volatility indicators such as VIX
and MOVE. The third is the uncertainty-duration channel captured by
persistence-related variables and persistence-by-stress interactions.

Formally, let $y_{t,h}$ denote the future volatility target at forecast
horizon $h$, let $\mathbf X_t$ denote a vector of conventional predictors,
and let $\mathbf Z_t$ denote a vector of persistence-based features. The
forecasting problem may be written as
\begin{equation}
  y_{t,h} \;=\; F\!\bigl(\mathbf X_t,\, \mathbf Z_t\bigr) + u_{t+h},
\end{equation}
where $F(\cdot)$ may represent either a linear or nonlinear forecasting
system. The substantive contribution of the framework lies less in the
specific estimator than in the construction of the persistence feature
space. The feature vector includes several dimensions of persistence
behavior:
\begin{equation}
\mathbf Z_t \;=\; \Bigl(
  \dhat_t,\,\Delta\dhat_t,\,\mathrm{Vol}(\dhat)_t,\,\mathrm{Trend}(\dhat)_t,\,
  H_t,\,\Delta H_t,\,
  \overline{d}_t,\,\sigma_d^t,\,\overline{d}_{s(i),t},\,
  \dhat_t\!\cdot\!\VIX_t,\,\dhat_t\!\cdot\!\mathrm{MOVE}_t
\Bigr).
\end{equation}
These variables summarize persistence dynamics at several levels. The
own-stock persistence estimate $\dhat_t$ captures local persistence
behavior. Cross-sectional aggregates $\overline d_t$ and $\sigma_d^t$
summarize system-wide persistence conditions, while sectoral averages
$\overline d_{s(i),t}$ capture industry-specific uncertainty propagation.
Interaction terms such as $\dhat_t\!\cdot\!\VIX_t$ allow the impact of
stress to depend on the persistence state of the market.

The forecasting architecture is organized as a layered ladder of models
designed to isolate the marginal contribution of different persistence
components. The baseline model is the standard HAR specification:
\begin{align}
  \textbf{Model A:}\quad
  &\log\RV^{(d)}_t,\;\log\RV^{(w)}_t,\;\log\RV^{(m)}_t,\;r_{t-1},\;|r_{t-1}|.
  \notag
\end{align}

The first augmentation introduces market-state variables:
\begin{align}
  \textbf{Model }A_1:\quad &A + \bigl(\VIX_t,\;\mathrm{MOVE}_t\bigr).
  \notag
\end{align}
This HAR-X structure measures the contribution of contemporaneous market
stress independently of persistence effects.

The second augmentation introduces own-stock persistence and roughness
features:
\begin{align}
  \textbf{Model }A_2:\quad
  &A + \bigl(\dhat_t,\;\Delta\dhat_t,\;\mathrm{Vol}(\dhat)_t,\;
  \mathrm{Trend}(\dhat)_t,\;H_t,\;\Delta H_t\bigr).
  \notag
\end{align}

The third and fourth augmentations isolate cross-sectional and sectoral
persistence effects:
\begin{align}
  \textbf{Model }A_3:\quad &A + \bigl(\overline{d}_t,\;\sigma_d^t\bigr),
  \notag\\
  \textbf{Model }A_4:\quad &A + \overline{d}_{s(i),t}.
  \notag
\end{align}

The fifth augmentation introduces persistence-by-stress interactions:
\begin{align}
  \textbf{Model }A_5:\quad
  &A + \bigl(\dhat_t,\;\VIX_t,\;\mathrm{MOVE}_t,\;
  \dhat_t\!\cdot\!\VIX_t,\;\dhat_t\!\cdot\!\mathrm{MOVE}_t\bigr).
  \notag
\end{align}

The complete structural forecasting system combines all persistence blocks
simultaneously:
\begin{align}
  \textbf{Model C:}\quad &A + A_2 + A_3 + A_4 + A_5.
  \notag
\end{align}

Finally, the same feature space is estimated using several machine-learning
procedures:
\begin{align}
  \textbf{Model D:}\quad
  &\text{Lasso, Ridge, Elastic Net, Random Forest, Gradient Boosting.}
  \notag
\end{align}

This layered design is important because it isolates the marginal
contribution of each persistence block relative to the HAR benchmark. The
framework therefore allows one to determine whether predictive improvements
arise primarily from market-state variables, own-stock persistence,
cross-sectional persistence, sectoral aggregation, or persistence-by-stress
interactions.

The architecture also allows evaluation of an important methodological
question: whether forecasting gains arise primarily from economically
structured feature design or from estimator flexibility itself. By
comparing linear persistence-based models with flexible machine-learning
estimators using the identical feature set, the framework isolates the
incremental contribution of nonlinear estimation methods separately from
the contribution of economically interpretable persistence features.

Several implementation safeguards are required because persistence-based
variables are estimated quantities rather than directly observed variables.
Rolling estimation windows are constructed to avoid look-ahead bias and to
preserve the forecasting information set. Feature standardization is
performed using training-sample information only. Hyperparameters are
selected through time-series cross-validation rather than random
cross-sectional shuffling. These precautions are particularly important
because volatility forecasting systems are highly sensitive to temporal
dependence structures.

Forecast evaluation also requires careful treatment because volatility
proxies themselves contain measurement noise. In addition to mean squared
error on the logarithmic realized-volatility scale, the framework therefore
employs the QLIKE loss function proposed by \citet{patton2011volatility},
which is more robust when realized-volatility proxies contain observational
error. For pairwise forecast comparisons, the paper uses the
Diebold--Mariano statistic of \citet{diebold1995comparing} with the
\citet{harvey1997testing} finite-sample correction (HLN), computed on the
cross-sectional mean loss differential per date and using a Newey--West HAC
variance with bandwidth $\lceil h/5\rceil-1$. Importantly, inference is
conducted using panel-aware cross-sectional aggregation rather than naive
pooled-cell independence assumptions. This distinction is critical because
ignoring within-date cross-sectional dependence substantially inflates
apparent statistical significance.

The forecasting structure developed here is therefore designed not merely
to maximize predictive accuracy, but to investigate whether persistence
measures behave like economically meaningful state variables. If
persistence-based features systematically improve forecasts during prolonged
stress episodes, particularly at longer horizons, then the interpretation
of persistence as an indicator of uncertainty duration gains substantial
empirical support.

As shown later in the empirical analysis, this is precisely what occurs.
Persistence-related variables provide their largest incremental forecasting
value during high-stress regimes and at longer forecast horizons, exactly
where the duration of uncertainty becomes economically most important.

\section{Data and Empirical Design}
\label{sec:data}

This section describes the data architecture, volatility construction
procedures, rolling estimation framework, and forecasting protocol used in
the empirical analysis. The central methodological principle is that
persistence and roughness measures must be estimated in a manner consistent
with the temporal structure of financial data while avoiding forward-looking
contamination and preserving comparability across forecasting models.

The empirical framework is designed around three objectives. First, the
data environment must be sufficiently long to capture multiple volatility
regimes, including calm periods, financial crises, and post-crisis
transitions. Second, the volatility proxies must be constructed carefully
enough to distinguish genuine persistence from measurement noise. Third,
the forecasting design must permit rigorous out-of-sample evaluation under
realistic financial forecasting conditions.

\subsection{Sample}

The primary dataset consists of common equities drawn from the S\&P~500
index with sufficiently long daily histories to support rolling estimation
of persistence and roughness measures. Daily open, high, low, close,
volume, and market-capitalization data are obtained from Bloomberg for an
initial universe of 125 securities. After imposing a 70\% data-coverage
requirement over the sample period, the final balanced panel contains
115 stocks observed over 6,136 trading days from November~29, 2001 through
April~21, 2026.\footnote{Accessed: 21 April 2026 , Bloomberg Terminal.}

The sample spans several major financial regimes, including the post-2001
recovery period, the credit expansion preceding the global financial
crisis, the global financial crisis itself, the European sovereign-debt
episode, the long low-volatility post-crisis expansion, the COVID shock,
and the post-COVID inflationary environment. The presence of these distinct
regimes is essential because the paper interprets persistence measures as
state-dependent indicators of uncertainty propagation.

In addition to stock-level data, the analysis incorporates several
market-wide variables intended to capture broad financial conditions. These
include the S\&P~500 index, Dow Jones Industrial Average, Nasdaq~100,
Russell~2000, the VIX implied-volatility index, the MOVE Treasury-volatility
index, Treasury-yield series, the 2s/10s yield-curve slope, and
investment-grade and high-yield CDX credit indices. Sector-level
aggregation is implemented using GICS Level-1 classifications together
with SPDR sector ETFs.

To supplement the daily data, the study also incorporates a smaller
high-frequency subsample consisting of five-minute observations for the
S\&P~500 and twenty liquid equities over the period October~2025 through
April~2026. These data are used primarily for robustness analysis involving
realized-volatility estimation and rough-volatility diagnostics rather than
as the primary forecasting sample.

The resulting dataset provides several important advantages. The long
sample horizon allows persistence measures to be evaluated across multiple
crisis periods and volatility states. The cross-sectional structure permits
construction of aggregate persistence measures and sectoral persistence
indices. The high-frequency subsample allows comparison between daily
volatility proxies and realized-volatility estimators.

\subsection{Volatility Proxies}

Daily logarithmic returns are computed from adjusted closing prices and
winsorized at the 0.1\% and 99.9\% percentiles within each stock in order
to reduce the influence of recording errors, corporate actions, and
isolated outlier observations. Because volatility measurement plays a
central role in persistence estimation, the choice of volatility proxy is
particularly important.

The primary volatility measure used throughout the paper is the Parkinson
range-based variance estimator,
\begin{equation}
  \RVPK_t \;=\; \frac{(\ln H_t - \ln L_t)^2}{4\ln 2},
  \label{eq:parkinson}
\end{equation}
where $H_t$ and $L_t$ denote the daily high and low prices. Unlike squared
close-to-close returns, the Parkinson estimator exploits intraday price
range information and therefore provides a more efficient volatility proxy
under continuous-diffusion assumptions. The Parkinson measure offers several
advantages for the present application. First, it is available consistently
over the entire 25-year sample period for all stocks in the panel. Second,
it substantially reduces measurement noise relative to squared returns.
Third, the estimator captures volatility clustering particularly clearly,
which is important for persistence estimation.

The empirical analysis also employs several alternative volatility proxies
for robustness purposes, including squared returns, absolute returns, and
realized variance constructed from high-frequency data where available.
These alternative measures help determine whether the persistence results
depend critically on the choice of volatility proxy.

For forecasting purposes, the volatility target for stock $i$ at horizon
$h$ is defined as the logarithm of future average realized variance,
\begin{equation}
  y_{i,t,h} \;=\; \log\!\Bigl(\tfrac{1}{h}\!\sum_{j=1}^{h}
  \RVPK_{i,t+j}\Bigr).
\end{equation}
Three forecast horizons are considered, $h\in\{1,5,22\}$, corresponding
approximately to daily, weekly, and monthly horizons. This multi-horizon
structure is important because the economic interpretation of persistence
predicts that persistence-related features should contribute most strongly
at longer horizons where uncertainty-duration effects become more relevant.

\subsection{Descriptive Statistics}

Table~\ref{tab:summary_stats} reports summary statistics for daily log
returns pooled across the 115-stock panel and broken down by GICS sector.
The pooled annualized mean return equals approximately 9.4\%, while the
pooled annualized standard deviation equals approximately 31.5\%.

More importantly, the distribution of returns displays substantial
non-normality. The pooled excess kurtosis exceeds 11, and the Jarque--Bera
statistic overwhelmingly rejects Gaussianity. A fitted Student-$t$
distribution implies degrees of freedom near 2.7, indicating extremely
heavy tails. Substantial cross-sectional heterogeneity is also present
across sectors. Information Technology and Financials display the largest
annualized volatility levels, whereas Utilities and Consumer Staples
display substantially lower volatility. Real Estate and Financials exhibit
especially heavy-tailed behavior, consistent with their sensitivity to
systemic financial conditions and leverage dynamics.

Table~\ref{tab:data_description} summarizes the composition of the panel
together with the distribution of market-wide state variables. The VIX
ranges from approximately 9 to above 82, while the MOVE index ranges from
approximately 37 to above 264. These substantial variations in market
stress are particularly important because the structural interpretation of
persistence predicts that persistence measures should rise systematically
during periods of elevated uncertainty.

The descriptive evidence therefore already suggests several key themes that
are explored later in the forecasting analysis. Financial returns are
highly non-Gaussian, volatility dynamics exhibit substantial persistence,
and market-wide stress conditions vary dramatically across the sample
period.
\FloatBarrier

\begin{table}[h!]
\centering
\caption{Summary Statistics for Daily Stock Returns}
\label{tab:summary_stats}
\small
\begin{tabular}{lcccccc}
\toprule
 & Mean & Std & Skewness & Kurtosis & Min & Max \\
 & (\% ann.) & (\% ann.) & & & (\%) & (\%) \\
\midrule
\multicolumn{7}{l}{\textbf{Panel A: Pooled Sample}} \\
All Stocks (N=115) & 9.38 & 31.48 & -0.13 & 11.64 & -25.60 & 26.73 \\
\midrule
\multicolumn{7}{l}{\textbf{Panel B: By GICS Sector}} \\
Communication Services (n=5) & 12.45 & 32.90 & -0.07 & 6.40 & -12.67 & 11.90 \\
Consumer Discretionary (n=9) & 12.53 & 30.26 & -0.05 & 5.91 & -11.45 & 10.87 \\
Consumer Staples (n=11) & 8.23 & 22.78 & -0.15 & 6.33 & -8.95 & 8.48 \\
Energy (n=5) & 8.59 & 30.83 & -0.24 & 5.82 & -11.72 & 11.25 \\
Financials (n=19) & 6.74 & 34.38 & -0.14 & 11.14 & -15.00 & 14.85 \\
Health Care (n=15) & 9.09 & 27.95 & -0.17 & 6.27 & -10.79 & 10.12 \\
Industrials (n=17) & 9.37 & 27.28 & -0.19 & 5.17 & -9.96 & 9.28 \\
Information Technology (n=18) & 11.41 & 36.16 & -0.13 & 5.22 & -13.04 & 12.17 \\
Materials (n=3) & 9.29 & 33.39 & -0.34 & 5.32 & -12.67 & 10.66 \\
Real Estate (n=7) & 11.80 & 34.15 & 0.11 & 11.17 & -13.87 & 15.37 \\
Utilities (n=6) & 6.35 & 22.39 & -0.12 & 7.65 & -8.89 & 9.57 \\
\bottomrule
\end{tabular}
\begin{tablenotes}\small
\item Notes: Sample period Nov 2001 -- Apr 2026, 115 S\&P 500 constituents that pass a 70\% coverage filter. Mean and Std are annualized; Skewness and Kurtosis are excess values. Returns are log returns winsorized at the 0.1\% and 99.9\% percentiles.
\end{tablenotes}
\end{table}
\FloatBarrier

\FloatBarrier

\begin{table}[h!]
\centering
\caption{Data Description}
\label{tab:data_description}
\small
\begin{tabular}{lcc}
\toprule
\multicolumn{3}{l}{\textbf{Panel A: Sample Composition by GICS Sector}} \\
Sector & N Stocks & \% of Sample \\
\midrule
Financials & 19 & 16.5\% \\
Information Technology & 18 & 15.7\% \\
Industrials & 17 & 14.8\% \\
Health Care & 15 & 13.0\% \\
Consumer Staples & 11 & 9.6\% \\
Consumer Discretionary & 9 & 7.8\% \\
Real Estate & 7 & 6.1\% \\
Utilities & 6 & 5.2\% \\
Communication Services & 5 & 4.3\% \\
Energy & 5 & 4.3\% \\
Materials & 3 & 2.6\% \\
\midrule
Total & 115 & 100.0\% \\
\bottomrule
\end{tabular}

\vspace{0.5cm}

\begin{tabular}{lccccc}
\toprule
\multicolumn{6}{l}{\textbf{Panel B: Market-Level Variables}} \\
Variable & Mean & Std & Min & Max & N \\
\midrule
SPX & 2398.64 & 1555.37 & 676.53 & 7126.06 & 6136 \\
INDU & 19990.01 & 11002.39 & 6547.05 & 50188.14 & 6136 \\
NDX & 6405.01 & 6378.05 & 804.64 & 26672.43 & 6136 \\
RTY & 1183.03 & 601.33 & 327.09 & 2792.96 & 6136 \\
VIX & 19.48 & 8.45 & 9.14 & 82.69 & 6136 \\
MOVE & 88.44 & 31.15 & 36.62 & 264.60 & 6136 \\
USGG3M & 1.71 & 1.83 & -0.14 & 5.51 & 6136 \\
USGG10YR & 3.15 & 1.16 & 0.51 & 5.43 & 6136 \\
USYC2Y10 & 110.42 & 96.62 & -108.71 & 291.03 & 6136 \\
CDX IG 5Y & 69.62 & 19.05 & 43.75 & 151.80 & 3674 \\
CDX HY 5Y & 104.59 & 4.19 & 86.25 & 110.41 & 3669 \\
\bottomrule
\end{tabular}
\begin{tablenotes}\small
\item Notes: Sample period Nov 2001 -- Apr 2026. Market-level variables sourced from Bloomberg.
\end{tablenotes}
\end{table}
\FloatBarrier

\subsection{Preliminary Diagnostics}

Before proceeding to rolling persistence estimation and forecasting
analysis, it is important to examine several basic empirical properties of
the data. The preliminary diagnostics confirm the presence of the central
stylized facts motivating the paper: volatility clustering, strong
nonlinear dependence in volatility measures, substantial heavy-tailed
behavior, cross-sectional comovement, and persistent conditional
heteroskedasticity.

\begin{figure}[H]
\centering
\includegraphics[width=\textwidth]{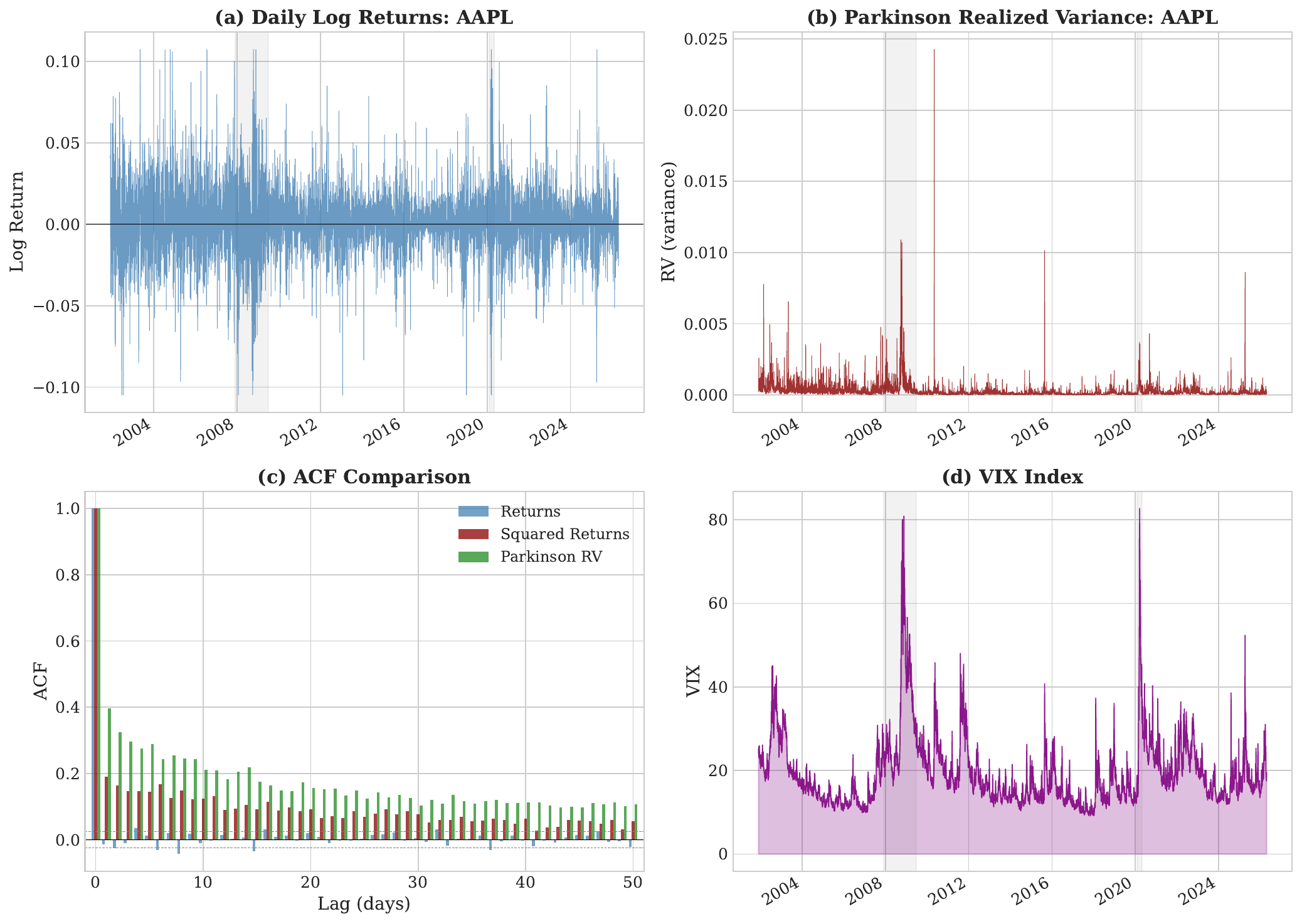}
\caption{Daily log returns and Parkinson range-based variance proxy for a
representative stock; ACF comparison; and the VIX time series.
Recession-shaded windows mark the global financial crisis and COVID.
\emph{Takeaway: volatility clusters visibly; squared returns and $\RVPK$
exhibit slow autocorrelation decay out to 50 lags while raw returns do
not.}}
\label{fig:data_overview}
\end{figure}

Figure~\ref{fig:data_overview} presents four preliminary diagnostics for a
representative stock, together with the VIX index. Panel~(a) displays
daily logarithmic returns for AAPL, while panel~(b) reports the
corresponding Parkinson range-based variance proxy. The contrast between
the two series immediately reveals one of the classical stylized facts of
financial econometrics. Returns themselves fluctuate around zero with
little visually detectable serial dependence, whereas volatility measures
display pronounced clustering. Large volatility episodes tend to be
followed by additional large volatility episodes, while calm periods tend
to persist as well.

The recession-shaded regions corresponding to the global financial crisis
and COVID are especially informative. During these periods, the Parkinson
variance proxy rises sharply and remains elevated for extended intervals
rather than reverting immediately to calm-state levels. This behavior is
precisely the type of prolonged uncertainty propagation that motivates the
interpretation of persistence measures as indicators of stress duration
rather than merely volatility magnitude.

Panel~(c) of Figure~\ref{fig:data_overview} compares autocorrelation
functions for raw returns, squared returns, and Parkinson realized
variance. The results are striking. Raw returns exhibit essentially no
significant serial dependence beyond the first few lags, consistent with
the approximate unpredictability of returns documented throughout the
financial-econometrics literature. By contrast, both squared returns and
the Parkinson variance proxy display autocorrelations that decay slowly and
remain significantly positive far beyond short horizons. The Parkinson
autocorrelations lie consistently above those for squared returns,
suggesting that the range-based estimator provides a less noisy volatility
proxy.

This distinction is important because persistence estimation is highly
sensitive to measurement error. Squared returns contain substantial noise
arising from isolated jumps and market microstructure effects. By exploiting
intraday high--low information, the Parkinson estimator substantially
improves the signal-to-noise ratio of volatility measurement.

Panel~(d) presents the VIX time series over the full sample period. The
VIX displays several dramatic spikes corresponding to major stress
episodes, particularly the global financial crisis and COVID. These broad
market stress indicators later play a central role in the
persistence-by-stress interaction framework developed in
Section~\ref{sec:ml}.

\begin{figure}[H]
\centering
\includegraphics[width=\textwidth]{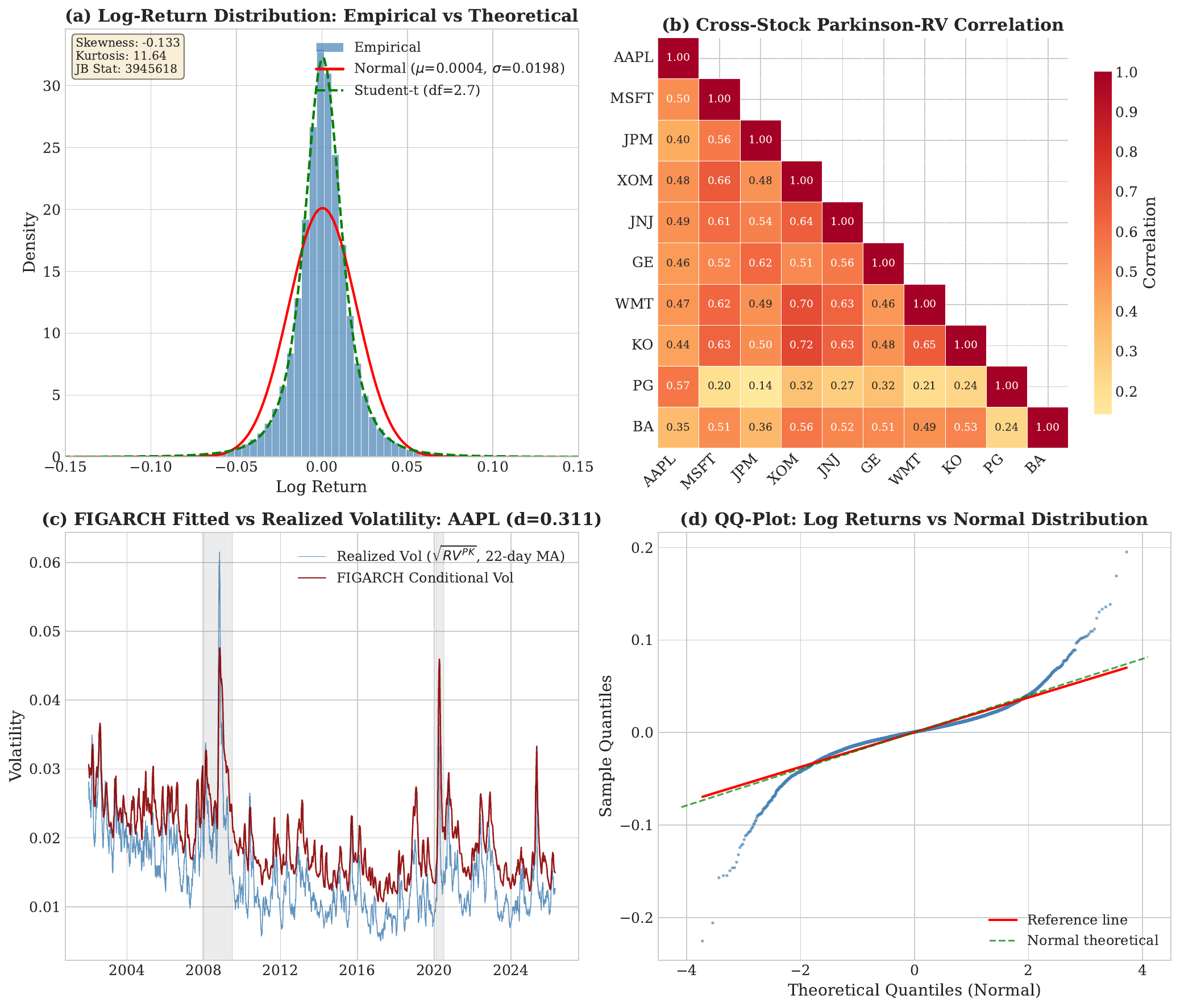}
\caption{Preliminary diagnostics. (a) Pooled log-return distribution with
Normal and Student-$t$ fits. (b) Cross-stock Parkinson-RV correlation
heatmap. (c) FIGARCH(1,$d$,1) conditional volatility versus realised
volatility for AAPL. (d) QQ-plot against a normal reference.
\emph{Takeaway: returns are heavily fat-tailed; FIGARCH reproduces
$\RVPK$-scale volatility well across the full sample.}}
\label{fig:diagnostics}
\end{figure}

Figure~\ref{fig:diagnostics} extends the diagnostic analysis in several
directions. Panel~(a) compares the empirical pooled distribution of daily
returns with fitted Gaussian and Student-$t$ distributions. The empirical
distribution displays substantially heavier tails than the Gaussian
benchmark, particularly in the extremes. The associated Jarque--Bera
statistic overwhelmingly rejects normality, while the QQ-plot in panel~(d)
displays the classical S-shaped departure from Gaussian behavior
characteristic of financial return distributions. Extreme observations
occur far more frequently than predicted by normal-diffusion models. This
heavy-tailed structure has important implications for volatility modeling
because persistence estimation can be distorted when tail behavior is
ignored.

Panel~(b) of Figure~\ref{fig:diagnostics} reports cross-stock Parkinson
variance correlations for ten representative equities, ranging from
$0.16$ (KO--AAPL) to $0.85$ (PG--CL). The heatmap reveals substantial
cross-sectional heterogeneity in volatility comovement, with particularly
large correlations among firms exposed to similar macroeconomic or sectoral
conditions. These cross-sectional relationships motivate the construction
of aggregate persistence variables later in the forecasting framework. If
uncertainty propagates systemically across sectors and balance sheets, then
cross-sectional persistence aggregates should contain information beyond
individual-stock volatility dynamics.

Panel~(c) compares FIGARCH conditional volatility estimates with rolling
Parkinson realized volatility for AAPL. The two series track each other
closely throughout the sample, including major stress periods. This result
provides an important preliminary validation of the fractional-volatility
framework. Table~\ref{tab:figarch_estimates} gives FIGARCH point estimates
for ten stocks with normal innovations: the mean estimate is
$\hat d = 0.329$ with values in $[0.237,\,0.461]$, all within the
stationary long-memory range $(0,\,\tfrac12)$ and consistent with the
semiparametric estimates we report below.

Taken together, the preliminary diagnostics establish several important
empirical conclusions. First, volatility measures exhibit strong persistence
while returns themselves remain approximately serially uncorrelated. Second,
financial returns are substantially heavy-tailed and strongly non-Gaussian.
Third, volatility dynamics display substantial cross-sectional comovement.
Fourth, long-memory volatility models provide empirically plausible
descriptions of observed volatility behavior. Most importantly, the
diagnostics strongly support the interpretation that volatility persistence
is not an isolated feature of a few securities or crisis episodes. Rather,
persistent volatility dynamics appear broadly throughout the cross-section
and evolve systematically across market states.

\FloatBarrier

\begin{table}[h!]
\centering
\caption{FIGARCH(1,d,1) Model Estimates}
\label{tab:figarch_estimates}
\small
\begin{tabular}{lcccccc}
\toprule
Stock & $\omega$ & $d$ & $\phi$ & $\beta$ & AIC & BIC \\
\midrule
AAPL & 0.2153 & 0.311 & 0.313 & 0.524 & 25240.0 & 25280.3 \\
MSFT & 0.2259 & 0.309 & 0.218 & 0.426 & 22734.1 & 22774.4 \\
JPM & 0.1213 & 0.461 & 0.182 & 0.510 & 23429.5 & 23469.8 \\
XOM & 0.0671 & 0.421 & 0.231 & 0.557 & 21309.8 & 21350.2 \\
JNJ & 0.0725 & 0.308 & 0.346 & 0.498 & 17233.5 & 17273.8 \\
GE & 0.0904 & 0.378 & 0.311 & 0.556 & 23508.0 & 23548.3 \\
WMT & 0.1606 & 0.244 & 0.378 & 0.483 & 19624.8 & 19665.1 \\
KO & 0.1433 & 0.273 & 0.000 & 0.175 & 17667.8 & 17708.1 \\
PG & 0.1897 & 0.237 & 0.000 & 0.116 & 17635.3 & 17675.6 \\
BA & 0.1723 & 0.346 & 0.322 & 0.568 & 24599.0 & 24639.4 \\
\midrule
\textbf{Mean} & -- & \textbf{0.329} & -- & -- & -- & -- \\
\bottomrule
\end{tabular}
\begin{tablenotes}\small
\item Notes: FIGARCH(1,d,1) with AR(1) mean and normal innovations. $d \in (0, 0.5)$ indicates stationary long memory in conditional variance.
\end{tablenotes}
\end{table}
\FloatBarrier

\subsection{Rolling Estimation Design}

Persistence and roughness measures are estimated within rolling windows so
that they behave as dynamic state variables rather than fixed full-sample
constants. This distinction is essential for the interpretation developed
throughout the paper. If persistence reflects evolving uncertainty
conditions, information diffusion, and regime-dependent stress propagation,
then persistence measures themselves must be allowed to vary over time.

For each stock $i$ and each end date $t$, the fractional integration
parameter $\dhat_{i,t}$ is estimated using a rolling 750-trading-day
window. Both the Geweke--Porter-Hudak log-periodogram estimator and the
local-Whittle semiparametric estimator are employed using bandwidth
$m = \lfloor T^{0.65}\rfloor$. The rolling estimates are updated every
five trading days, producing weekly time series of
$\dhat_{i,t}^{GPH}$, $\dhat_{i,t}^{LW}$, and $H_{i,t}$, where $H_{i,t}$
denotes the rolling Hurst exponent estimated from logarithmic Parkinson
realized variance.

The choice of weekly updating frequency is intentional. Daily updating
would produce extremely noisy persistence estimates dominated by local
estimation variation rather than economically meaningful state changes.
Monthly updating, on the other hand, would smooth away important regime
transitions. Weekly updating provides an intermediate compromise that
remains responsive to changing market conditions while preserving
reasonable statistical stability.

The 750-day estimation window is similarly motivated by the tradeoff
between statistical precision and state sensitivity. Persistence estimation
requires relatively long windows because long-memory parameters are
fundamentally low-frequency objects. Very short windows generate unstable
estimates with large sampling variance. Extremely long windows, however,
blur structural regime changes and reduce responsiveness to evolving market
conditions. The selected window length therefore balances these competing
considerations.

The rolling framework transforms persistence from a static full-sample
coefficient into a genuinely time-varying market-state variable. This
transformation is central to the economic interpretation of the paper.
Under the rolling design, persistence measures may rise during prolonged
stress episodes and decline during calm periods, allowing direct analysis
of uncertainty-duration dynamics.

From these rolling estimates, the paper constructs the persistence feature
vector introduced in Section~\ref{sec:ml}. The feature set includes several
categories of variables. The first category consists of own-stock
persistence dynamics,
\begin{equation}
  \dhat_{i,t},\;\Delta\dhat_{i,t},\;
  \mathrm{Vol}(\dhat)_{i,t},\;\mathrm{Trend}(\dhat)_{i,t},\;
  H_{i,t},\;\Delta H_{i,t},
\end{equation}
summarizing the level, evolution, variability, and trend structure of
persistence and roughness at the individual-stock level. The second
category consists of cross-sectional persistence aggregates,
\begin{equation}
  \overline d_t,\;\sigma_d^t,\;\mathrm{Skew}_t(d),\;\mathrm{Kurt}_t(d),
\end{equation}
which summarize system-wide persistence conditions. The cross-sectional
mean $\overline d_t$ is especially important because the paper interprets
it as a broad indicator of the duration of unresolved uncertainty in
financial markets. The third category consists of sectoral persistence
measures, $\overline d_{s(i),t}$, where the average is computed across
stocks belonging to the same GICS sector as stock $i$, intended to capture
industry-specific stress propagation and heterogeneous information
environments.

Additional features include threshold indicators $\mathbf 1_{\{\dhat_{i,t}
> \tau\}}$ together with interaction terms involving VIX, MOVE, and
liquidity conditions. These interaction variables allow the forecasting
impact of persistence to depend explicitly on market stress states.

Table~\ref{tab:features} reports pooled descriptive statistics for the
resulting feature set. Several patterns are immediately noteworthy. First,
the local-Whittle estimates are systematically larger than the GPH
estimates, consistent with earlier long-memory studies. Second, the Hurst
estimates remain uniformly well below $\tfrac12$, confirming rough-volatility
behavior. Third, the cross-sectional mean persistence measure varies
substantially over time, supporting its interpretation as a dynamic state
variable rather than a stable structural constant. The persistence-by-VIX
interaction variable becomes very large during crisis periods, suggesting
that persistence and market stress reinforce one another during prolonged
uncertainty episodes.

An important feature of the rolling design is that all variables are
constructed strictly using information available at time $t$. This prevents
forward-looking contamination and ensures that the forecasting exercise
represents a genuine out-of-sample evaluation rather than a pseudo-exercise
using future information implicitly embedded in the feature set.

The rolling framework also provides a natural bridge between econometric
estimation and economic interpretation. Under the structural interpretation
developed earlier, persistence should rise during periods in which
uncertainty dissipates slowly and should decline during periods in which
shocks are absorbed rapidly. The rolling estimates allow this prediction to
be tested directly in the empirical analysis.

Moreover, because the rolling framework generates persistence measures for
each stock, sector, and date simultaneously, it becomes possible to study
not only average persistence but also the cross-sectional dispersion and
heterogeneity of persistence conditions throughout the financial system.
The resulting persistence-state panel therefore forms the core empirical
object underlying the forecasting analysis developed in the next section.
\FloatBarrier

\begin{table}[htbp]
\centering
\caption{Persistence Feature Vector: Definitions and Pooled Statistics}
\label{tab:features}
\small
\begin{tabular}{llcccc}
\toprule
Category & Feature & Mean & Std & 1\% & 99\% \\
\midrule
\multicolumn{6}{l}{\textbf{LRD}} \\
  & $\hat d_{GPH,t}$ (rolling) & 0.1731 & 0.0961 & 0.0008 & 0.4154 \\
  & $\hat d_{LW,t}$ (rolling) & 0.3358 & 0.1738 & 0.0265 & 0.8278 \\
\midrule
\multicolumn{6}{l}{\textbf{Roughness}} \\
  & $H_t$ (rolling) & 0.0620 & 0.0275 & 0.0196 & 0.1303 \\
\midrule
\multicolumn{6}{l}{\textbf{Memory dyn.}} \\
  & $\Delta\hat d_t$ & -0.0001 & 0.0150 & -0.0478 & 0.0478 \\
  & $\mathrm{Vol}(\hat d)_t$ & 0.0067 & 0.0112 & 0.0000 & 0.0587 \\
  & $\mathrm{Trend}(\hat d)_t$ & -0.0001 & 0.0080 & -0.0266 & 0.0252 \\
\midrule
\multicolumn{6}{l}{\textbf{HAR}} \\
  & $RV^d$ & 0.0003 & 0.0012 & 0.0000 & 0.0027 \\
  & $RV^w$ & 0.0003 & 0.0010 & 0.0000 & 0.0027 \\
  & $RV^m$ & 0.0003 & 0.0008 & 0.0000 & 0.0025 \\
\midrule
\multicolumn{6}{l}{\textbf{Sector aggregate}} \\
  & $\bar d_{s(i),t}$ & 0.1730 & 0.0803 & 0.0446 & 0.3362 \\
\midrule
\multicolumn{6}{l}{\textbf{Cross-sectional}} \\
  & $\bar d_t$ (cross-sectional mean) & 0.1731 & 0.0746 & 0.0700 & 0.3029 \\
  & $\sigma_d^t$ (cross-sectional std) & 0.0596 & 0.0118 & 0.0335 & 0.0838 \\
  & Skew of $d_t$ across stocks & 0.0165 & 0.4559 & -1.1219 & 0.8546 \\
  & Kurt of $d_t$ across stocks & 0.3385 & 0.8753 & -0.7179 & 2.8986 \\
  & $P(\hat d > 0.30)$ & 0.1169 & 0.1747 & 0.0000 & 0.5217 \\
  & Range of $\hat d$ across stocks & 0.3077 & 0.0695 & 0.1905 & 0.4992 \\
\midrule
\multicolumn{6}{l}{\textbf{Market}} \\
  & VIX & 19.0494 & 8.3693 & 9.8977 & 53.7306 \\
  & MOVE & 84.6779 & 30.7937 & 44.3771 & 188.2890 \\
  & USYC2Y10 & 96.3550 & 94.2156 & -82.3963 & 280.2324 \\
\midrule
\multicolumn{6}{l}{\textbf{Interaction}} \\
  & $\hat d_t \cdot \mathrm{VIX}_t$ & 3.6108 & 3.1447 & 0.0125 & 14.7306 \\
  & $\hat d_t \cdot \mathrm{MOVE}_t$ & 15.3338 & 11.6301 & 0.0536 & 54.1197 \\
  & $\hat d_t / \mathrm{Liq}_t$ & 0.0000 & 0.0000 & 0.0000 & 0.0000 \\
\bottomrule
\end{tabular}
\begin{tablenotes}\small
\item Notes: Statistics are pooled across all stocks and dates in the rolling-estimation panel. $\hat d_t$ values are taken on a weekly stride; HAR components are constructed from Parkinson realised variance. Sector aggregates use GICS level-1; threshold indicators and interaction terms (with VIX, MOVE, and inverse dollar-volume liquidity) use the GPH estimate of $\hat d_t$.
\end{tablenotes}
\end{table}
\FloatBarrier

\subsection{Out-of-Sample Forecasting Protocol}

The forecast evaluation is strictly walk-forward. We use the first 40\%
of the sample dates (sample dates 1 through 431, ending mid-2013) as the
initial training window; thereafter we re-fit each model and produce a
forecast at every subsequent sample date. For the linear ladder
(Models~$A,A_1,A_2,A_3,A_4,A_5,C$) we re-fit by ordinary least squares
at every step (expanding window). For Model~$D$ we re-fit every twenty
sample steps to keep the computational burden tractable for the tree-based
estimators; results are not sensitive to this stride. All feature
standardization, hyperparameter selection, and model selection are
performed on training-window information only. Both the realized target
and the forecast are recorded as $(T_{\mathrm{eval}}\!\times\!N)$ panels,
which then feed the loss functions, the Diebold--Mariano test, and the
regime/sector splits in Section~\ref{sec:results}.

\section{Numerical Analysis and Forecasting Results}
\label{sec:results}

This section presents the empirical results of the paper. The analysis
proceeds in several stages. First, we examine full-sample estimates of
long-memory and roughness parameters across the cross-section of equities.
Second, we study the dynamics of rolling persistence measures across market
regimes. Third, we evaluate the forecasting performance of the layered
forecasting framework developed in Section~\ref{sec:ml}. Finally, we
investigate robustness across sectors, volatility states, liquidity
conditions, and alternative estimation procedures.

The central empirical question is whether persistence measures behave like
economically meaningful state variables rather than purely statistical
nuisance parameters. Under the interpretation developed earlier, persistence
measures should rise during sustained stress episodes, co-move with broad
uncertainty indicators, and contribute forecasting information particularly
during periods in which uncertainty resolves slowly.

\subsection{Long-Memory and Roughness Estimates}

We begin with full-sample cross-sectional estimates of persistence and
roughness. The results strongly confirm the simultaneous presence of
long-memory behavior and rough volatility across the equity panel.

The first important finding is that raw returns themselves display
essentially no evidence of long memory. The cross-sectional mean GPH
estimate for returns is approximately $\dhat^{GPH}_{\mathrm{returns}}
= -0.011$, with almost no stocks exhibiting statistical significance at
conventional levels. This result is consistent with the approximate
unpredictability of financial returns documented throughout the empirical
asset-pricing literature.

The situation changes dramatically once one examines volatility measures.
For the Parkinson range-based variance proxy, the cross-sectional mean GPH
estimate equals approximately $\dhat^{GPH}_{\mathrm{RV}} = 0.226$, while
the corresponding local-Whittle estimate equals approximately
$\dhat^{LW}_{\mathrm{RV}} = 0.440$. Moreover, between 98\% and 100\% of
stocks exhibit statistically significant persistence estimates depending on
the estimator employed. These results provide strong evidence that
long-memory behavior is a broad characteristic of volatility dynamics
rather than an isolated feature of a few securities.

The rough-volatility results are equally striking. The cross-sectional
mean Hurst exponent estimated from logarithmic Parkinson realized variance
equals approximately $H = 0.063$, with every stock in the panel satisfying
$H < \tfrac12$. This result implies that volatility is universally rough
at short horizons throughout the equity panel. The cross-sectional
dispersion of Hurst estimates is also remarkably small, suggesting that
rough-volatility behavior is highly stable across sectors and firms.

Table~\ref{tab:lrd_estimates} reports these estimates separately by GICS
sector. Financials and Real Estate exhibit among the largest persistence
estimates, consistent with their exposure to systemic stress propagation,
leverage conditions, and liquidity cycles. Utilities and Consumer Staples
display somewhat smaller persistence estimates, although long-memory
behavior remains substantial throughout all sectors. An important feature
of the results is the remarkable stability of persistence estimates across
the cross-section: the standard deviation of persistence estimates across
stocks remains relatively modest, indicating that long-memory volatility
behavior is fundamentally a panel-wide phenomenon rather than the
consequence of a few extreme outliers.
\FloatBarrier

\begin{table}[htbp]
\centering
\caption{Long-Range Dependence and Roughness Estimates by Sector}
\label{tab:lrd_estimates}
\small
\begin{tabular}{lccccccc}
\toprule
 & \multicolumn{2}{c}{Returns} & \multicolumn{3}{c}{Parkinson RV} & Roughness & \\
\cmidrule(lr){2-3} \cmidrule(lr){4-6}
Sector & $\bar d_{GPH}$ & $\bar d_{LW}$ & $\bar d_{GPH}$ & $\bar d_{LW}$ & \% Sig & $\bar H$ & N \\
\midrule
Communication Services & 0.009 & 0.002 & 0.239 & 0.477 & 100\% & 0.076 & 5 \\
Consumer Discretionary & -0.009 & -0.028 & 0.227 & 0.444 & 100\% & 0.056 & 9 \\
Consumer Staples & -0.018 & -0.041 & 0.192 & 0.381 & 91\% & 0.059 & 11 \\
Energy & -0.004 & -0.030 & 0.208 & 0.439 & 100\% & 0.067 & 5 \\
Financials & -0.013 & -0.033 & 0.245 & 0.461 & 100\% & 0.071 & 19 \\
Health Care & -0.017 & -0.030 & 0.215 & 0.419 & 100\% & 0.058 & 15 \\
Industrials & -0.005 & -0.015 & 0.222 & 0.437 & 94\% & 0.060 & 17 \\
Information Technology & -0.007 & -0.017 & 0.234 & 0.455 & 100\% & 0.061 & 18 \\
Materials & -0.010 & -0.017 & 0.232 & 0.448 & 100\% & 0.061 & 3 \\
Real Estate & -0.005 & -0.037 & 0.255 & 0.486 & 100\% & 0.069 & 7 \\
Utilities & -0.032 & -0.066 & 0.200 & 0.404 & 100\% & 0.062 & 6 \\
\midrule
\textbf{Overall} & -0.011 & -0.028 & 0.226 & 0.440 & 98\% & 0.063 & 115 \\
\bottomrule
\end{tabular}
\begin{tablenotes}\small
\item Notes: $d$ is the fractional differencing parameter. GPH is the Geweke--Porter--Hudak log-periodogram estimator (bandwidth $m = T^{0.65}$); LW is the local-Whittle semiparametric estimator. Returns are daily log returns; Parkinson RV is the range-based realized variance from daily H/L. $H$ is the Hurst exponent of $\log\mathrm{RV}^{PK}$ estimated by the scaling of the $q=2$ moment of increments over lags $\{1,2,3,5,8,13,21\}$. $H<0.5$ indicates rough behaviour. \% Sig reports the share of stocks whose volatility $d_{GPH}$ is significant at 5\%.
\end{tablenotes}
\end{table}
\FloatBarrier

\begin{figure}[H]
\centering
\includegraphics[width=\textwidth]{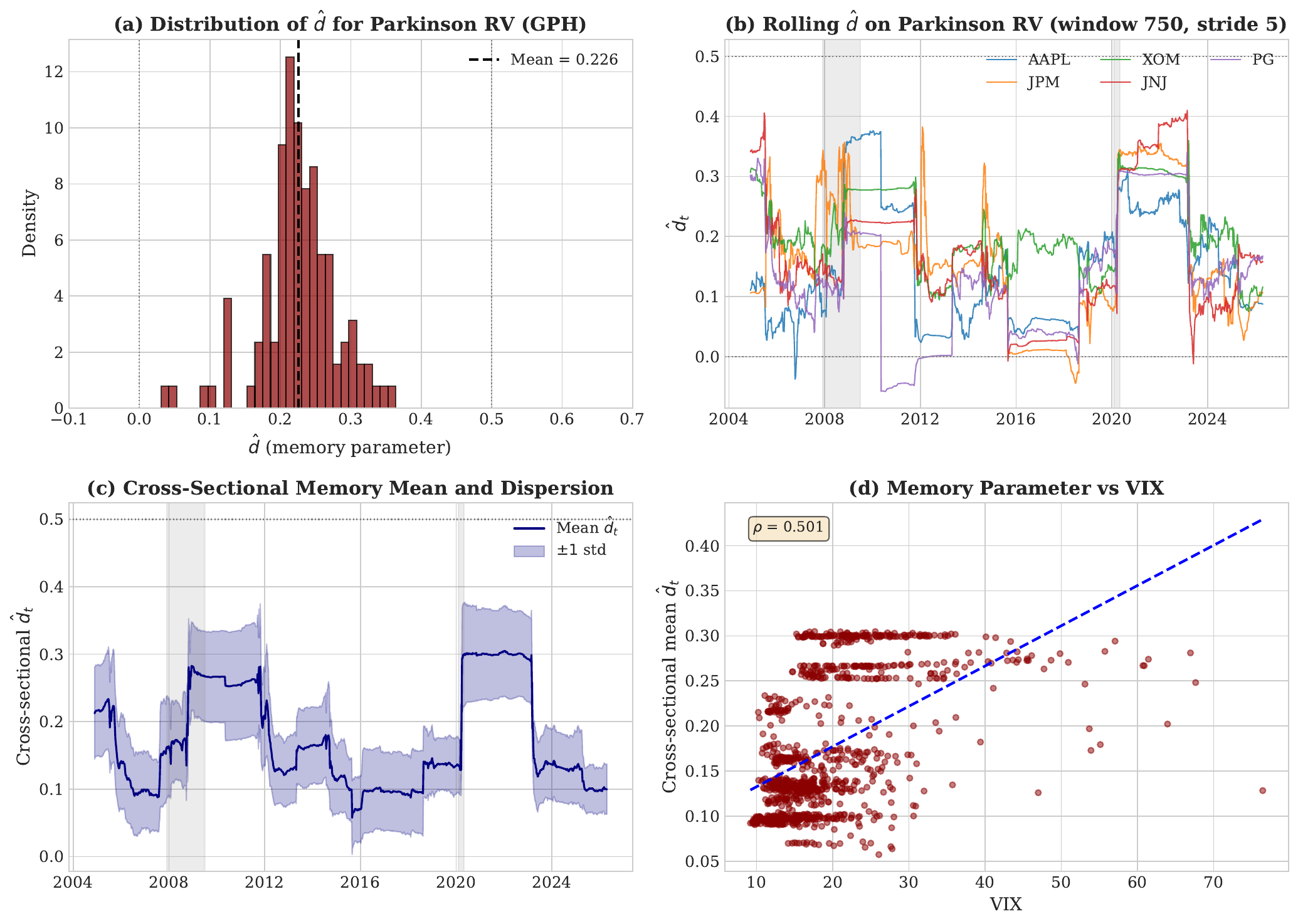}
\caption{Rolling persistence estimates. (a)~Cross-sectional distribution
of full-sample GPH $\dhat$ on Parkinson range-based variance proxy.
(b)~Rolling $\dhat^{GPH}$ for AAPL, JPM, XOM, JNJ, PG.
(c)~Cross-sectional mean $\overline d_t$ with $\pm 1$ standard deviation
band. (d)~$\overline d_t$ versus VIX, with linear fit and correlation.
\emph{Takeaway: the cross-sectional mean memory parameter rises sharply
in the GFC and COVID and tracks the VIX with correlation $\rho=+0.50$,
consistent with its interpretation as a duration-of-stress indicator.}}
\label{fig:lrd_estimates}
\end{figure}

Figure~\ref{fig:lrd_estimates} reports four time-series and cross-sectional
views of the rolling estimation. Panel~(a) reports the distribution of full-sample GPH estimates across all 115 stocks. The distribution is tightly concentrated around the interval $0.15 \leq \dhat \leq 0.30 $, with
relatively few extreme values.

Panel~(b) reports rolling $\dhat^{GPH}$ for five representative stocks
(AAPL, JPM, XOM, JNJ, PG). The most striking feature is the broad upward
shift in persistence during major crisis episodes, particularly during the
global financial crisis and COVID.

Panel~(c) presents perhaps the most important empirical figure in the
paper: the cross-sectional mean persistence state $\overline d_t =
N^{-1}\sum_i \dhat_{i,t}$, together with a one-standard-deviation band.
The dynamics of this aggregate persistence state are highly informative.
During the calm 2013--2014 period, the cross-sectional mean persistence
measure remains near approximately $\overline d_t \approx 0.154$. During
the global financial crisis, the cross-sectional mean rises to
approximately $\overline d_t \approx 0.259$, representing an increase of
approximately 68\%. During COVID, the persistence state rises even further
to approximately $\overline d_t \approx 0.287$, an increase of
approximately 86\% relative to the calm-state baseline.

These increases are economically large and strongly consistent with the
interpretation of persistence as an indicator of uncertainty duration.
Financial crises are not merely periods of elevated volatility. They are
periods in which uncertainty remains unresolved for extended intervals. The
persistence state appears to measure precisely this dimension.

Panel~(d) of Figure~\ref{fig:lrd_estimates} examines the relationship
between the cross-sectional persistence state and the VIX. The estimated
contemporaneous correlation equals approximately
$\rho(\overline d_t, \VIX_t) = 0.50$. This result is especially important
because it demonstrates that persistence and contemporaneous market stress
are closely related but not identical. The VIX primarily measures the
intensity of current uncertainty, whereas persistence appears to measure
the expected duration of uncertainty propagation. The distinction becomes
clear during prolonged crisis periods. Elevated volatility may persist
even after the initial shock has occurred because uncertainty continues to
propagate gradually through balance sheets, funding conditions, leverage
adjustments, and institutional portfolios.

Taken together, these findings provide direct empirical support for the
reduced-form interpretation of Section~\ref{sec:structural}. The rolling
persistence state behaves like a dynamic indicator of how slowly the
financial system is absorbing uncertainty. It rises during broad-based
stress episodes and declines gradually as uncertainty dissipates. The next
subsection investigates whether these persistence states also contain
economically meaningful forecasting information.

\subsection{Out-of-Sample Forecast Comparison}

We now evaluate the forecasting performance of the layered forecasting
framework introduced in Section~\ref{sec:ml}. The analysis focuses on
whether persistence-related features improve volatility forecasts beyond
standard HAR and HAR-X benchmarks and, more importantly, whether such
improvements occur precisely in the market states predicted by the
interpretation developed earlier.

\begin{figure}[H]
\centering
\includegraphics[width=\textwidth]{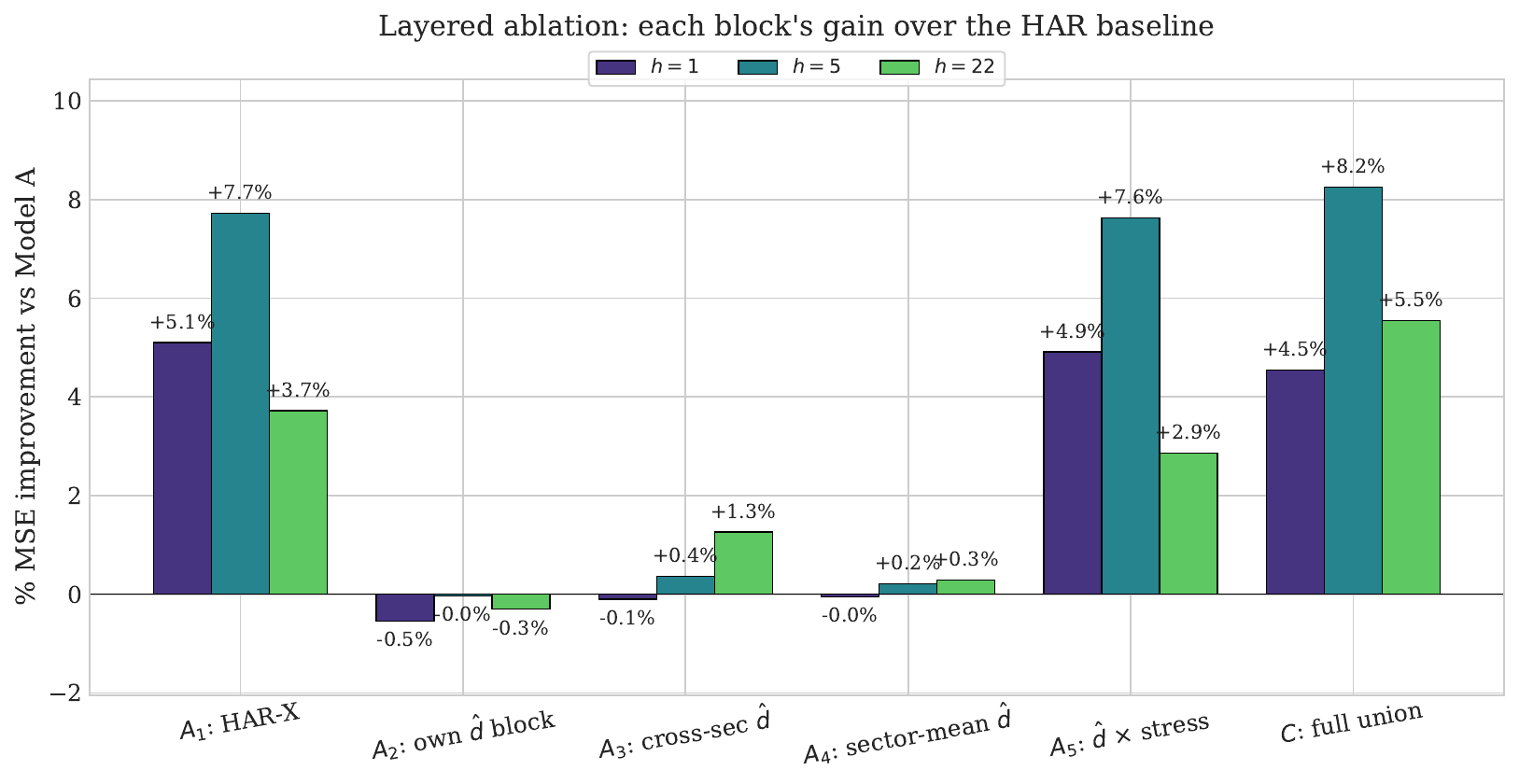}
\caption{Layered ablation of forecast gains relative to the HAR baseline.
Bars report the percentage MSE improvement of each model block over
Model~A at forecast horizons $h\in\{1,5,22\}$.}
\label{fig:layered_ablation}
\end{figure}

Table~\ref{tab:model_comparison} reports the main out-of-sample forecasting
results across the full forecasting ladder
$A\to A_1\to A_2\to A_3\to A_4\to A_5\to C\to D$, evaluated at the
daily, weekly, and monthly forecast horizons $h\in\{1,5,22\}$ and Figure~\ref{fig:layered_ablation} summarizes the layered ablation visually. The figure makes clear that most of the pooled forecasting improvement comes from market-state variables, while the additional persistence-related gains become most visible at the monthly horizon. Forecast
performance is measured primarily through mean squared error on the
logarithmic Parkinson realized-volatility scale together with the QLIKE
loss function. Statistical significance is evaluated using the
Harvey--Leybourne--Newbold finite-sample-corrected Diebold--Mariano
statistic computed under panel-aware cross-sectional aggregation.

The HAR baseline (Model~A) produces mean squared errors approximately
equal to $0.669$, $0.363$, $0.269$ for the daily, weekly, and monthly
horizons respectively. We note in passing that the iid pooled-cell DM
statistic ignores within-date cross-sectional dependence and is inflated
by roughly a factor of six relative to the panel-aware HLN-corrected
version (e.g., $+24.0$ versus $+3.87$ for Model~C at $h=5$); we report
the HLN values throughout.


\begin{table}[htbp]
\centering
\caption{Out-of-Sample Forecast Comparison: Layered Ladder $A \to A_1\dots A_5 \to C \to D$, Horizons $h \in \{1,5,22\}$}
\label{tab:model_comparison}
\small
\begin{tabular}{llcccc}
\toprule
Model & $h$ & MSE($\log RV^{PK}$) & QLIKE & \%$\Delta$ vs A & HLN DM-$t$ vs A \\
\midrule
\textbf{A} &  1 & 0.6686 & -7.5872 & +0.00\% & -- \\
 &  5 & 0.3634 & -7.6731 & +0.00\% & -- \\
 & 22 & 0.2689 & -7.6121 & +0.00\% & -- \\
\midrule
\textbf{A1} &  1 & 0.6345 & -7.6340 & +5.10\% & $+5.34^{***}$ \\
 &  5 & 0.3353 & -7.6949 & +7.72\% & $+5.31^{***}$ \\
 & 22 & 0.2589 & -7.6156 & +3.72\% & $+1.98^{**}$ \\
\midrule
\textbf{A2} &  1 & 0.6722 & -7.5780 & -0.54\% & $-3.06^{***}$ \\
 &  5 & 0.3635 & -7.6712 & -0.03\% & $-0.07$ \\
 & 22 & 0.2697 & -7.6083 & -0.30\% & $-0.41$ \\
\midrule
\textbf{A3} &  1 & 0.6692 & -7.5918 & -0.10\% & $-0.58$ \\
 &  5 & 0.3621 & -7.6755 & +0.37\% & $+1.05$ \\
 & 22 & 0.2655 & -7.6179 & +1.26\% & $+1.41$ \\
\midrule
\textbf{A4} &  1 & 0.6689 & -7.5843 & -0.05\% & $-0.45$ \\
 &  5 & 0.3627 & -7.6717 & +0.21\% & $+0.80$ \\
 & 22 & 0.2681 & -7.6108 & +0.29\% & $+0.47$ \\
\midrule
\textbf{A5} &  1 & 0.6358 & -7.6442 & +4.91\% & $+4.02^{***}$ \\
 &  5 & 0.3357 & -7.6970 & +7.63\% & $+4.18^{***}$ \\
 & 22 & 0.2612 & -7.6170 & +2.87\% & $+1.38$ \\
\midrule
\textbf{C} &  1 & 0.6382 & -7.6422 & +4.55\% & $+3.15^{***}$ \\
 &  5 & 0.3334 & -7.6992 & +8.24\% & $+3.87^{***}$ \\
 & 22 & 0.2540 & -7.6258 & +5.54\% & $+2.33^{**}$ \\
\midrule
\textbf{D\_lasso} &  1 & 0.6403 & -7.6316 & +4.23\% & $+3.92^{***}$ \\
 &  5 & 0.3381 & -7.6940 & +6.97\% & $+4.16^{***}$ \\
 & 22 & 0.2680 & -7.6113 & +0.33\% & $+0.20$ \\
\midrule
\textbf{D\_ridge} &  1 & 0.6452 & -7.6301 & +3.50\% & $+3.28^{***}$ \\
 &  5 & 0.3418 & -7.6931 & +5.95\% & $+3.60^{***}$ \\
 & 22 & 0.2698 & -7.6121 & -0.32\% & $-0.08$ \\
\midrule
\textbf{D\_en} &  1 & 0.6405 & -7.6305 & +4.20\% & $+3.97^{***}$ \\
 &  5 & 0.3382 & -7.6939 & +6.93\% & $+4.23^{***}$ \\
 & 22 & 0.2677 & -7.6117 & +0.44\% & $+0.25$ \\
\midrule
\textbf{D\_rf} &  1 & 0.6658 & -7.6177 & +0.42\% & $+0.41$ \\
 &  5 & 0.3553 & -7.6824 & +2.22\% & $+1.62$ \\
 & 22 & 0.2831 & -7.6002 & -5.29\% & $-2.24^{**}$ \\
\midrule
\textbf{D\_gbm} &  1 & 0.7584 & -7.5710 & -13.43\% & $-8.58^{***}$ \\
 &  5 & 0.3993 & -7.6656 & -9.87\% & $-4.27^{***}$ \\
 & 22 & 0.3206 & -7.5854 & -19.22\% & $-5.53^{***}$ \\
\bottomrule
\end{tabular}
\begin{tablenotes}\small
\item Notes: pooled across all stocks and out-of-sample dates (post 40\% warm-up). MSE on $\log RV^{PK}$ scale (range-based variance proxy); QLIKE on the variance scale. HLN DM-$t$ is the Harvey-Leybourne-Newbold finite-sample-corrected Diebold-Mariano statistic, computed on the cross-sectional mean loss differential per date with Newey-West HAC variance (bandwidth tied to $\lceil h / 5 \rceil - 1$ to handle overlapping multi-step forecasts on a weekly sample stride) and Student-$t(T-1)$ reference. Positive values indicate the model beats Model A. Significance: $^{*}$ $p<0.10$, $^{**}$ $p<0.05$, $^{***}$ $p<0.01$. Layered ladder: $A$ (HAR core), $A_1$ ($+$ VIX, MOVE; HAR-X), $A_2$ ($+$ own-stock persistence block), $A_3$ ($+$ cross-sectional mean and dispersion of $\hat d$), $A_4$ ($+$ sector-mean $\hat d$), $A_5$ ($+$ $\hat d$, VIX, MOVE, $\hat d \times$ VIX, $\hat d \times$ MOVE), $C$ (full union), $D$ (predictors of $C$ estimated by shrinkage / tree-based ML).
\end{tablenotes}
\end{table}

The first major result concerns the impact of market-state variables.
Adding only the VIX and MOVE controls (Model~$A_1$, the HAR-X
specification) produces substantial forecasting improvements at all
horizons: $+5.10\%$, $+7.72\%$, $+3.72\%$. All improvements are
statistically significant at conventional levels. This result is
economically intuitive. Implied-volatility measures summarize broad market
stress conditions and therefore contain substantial forecasting information
regarding near-term realized volatility.

The second result is considerably more interesting. Adding only own-stock
persistence and roughness variables (Model~$A_2$) produces essentially no
forecasting improvement: $-0.54\%$, $-0.03\%$, $-0.30\%$. In several
cases the forecasting performance slightly deteriorates relative to the
HAR benchmark itself. This finding is important because it clarifies the
role of persistence. Own-stock persistence by itself contributes little
incremental information beyond the multi-horizon HAR structure. The daily,
weekly, and monthly HAR components already capture much of the predictable
persistence structure within individual volatility series.

The results change once persistence is aggregated cross-sectionally.
Adding only the cross-sectional persistence aggregates (Model~$A_3$)
produces small but positive forecasting improvements: $-0.10\%$, $+0.37\%$,
$+1.26\%$. Similarly, sectoral persistence aggregates (Model~$A_4$) produce
modest positive gains, particularly at the monthly horizon. Although these
gains remain relatively small in isolation, their pattern is highly
informative. The improvements become larger precisely at longer horizons
where uncertainty-duration effects should matter most.

The persistence-by-stress interaction specification (Model~$A_5$) produces
substantially larger improvements: $+4.91\%$, $+7.63\%$, $+2.87\%$. At
the daily and weekly horizons, these gains closely resemble those obtained
by the HAR-X model. At the monthly horizon, however, the interaction
structure begins to diverge from the pure market-state specification.

The full structural model (Model~C), which combines all persistence blocks
simultaneously, produces the strongest overall forecasting performance:
$+4.55\%$, $+8.24\%$, $+5.54\%$. The associated HLN-corrected
Diebold--Mariano statistics equal approximately $+3.15$, $+3.87$, $+2.33$,
implying strong statistical significance at all horizons.

Several aspects of these results deserve emphasis.

First, most of the overall forecasting gain relative to the HAR baseline is
captured by adding market-state variables alone through HAR-X. This result
is not surprising because implied-volatility indices already summarize
substantial information about current market conditions.

Second, persistence features contribute their largest incremental gains at
the monthly horizon. Relative to HAR-X, the full structural model adds
approximately 1.8 additional percentage points of forecasting improvement
at $h=22$. This pattern is exactly what the structural interpretation
predicts. Persistence measures are intended to capture the expected duration
of uncertainty propagation rather than merely the level of current
volatility. Duration effects should therefore matter more at longer horizons
than at very short horizons.

\begin{figure}[H]
\centering
\includegraphics[width=\textwidth]{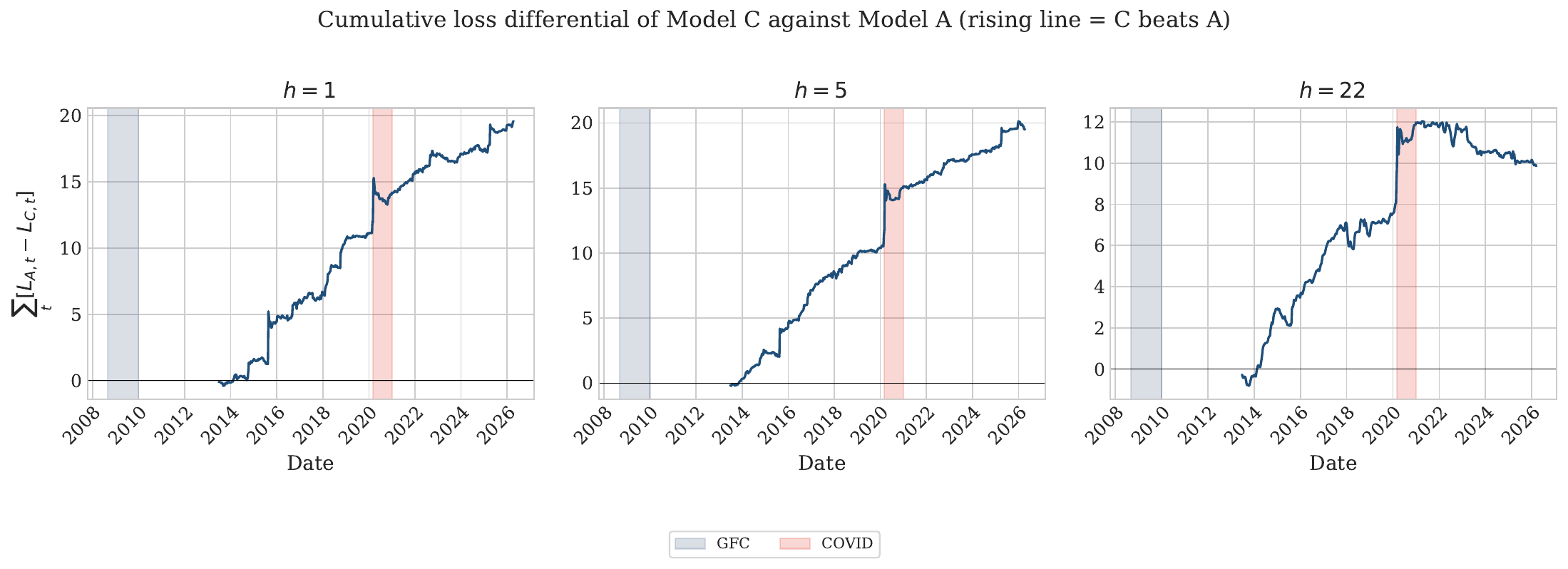}
\caption{Cumulative loss differential $\sum_t (L_{A,t}-L_{C,t})$ across
out-of-sample dates, separately for $h\in\{1,5,22\}$. A rising line
indicates Model~$C$ beats Model~$A$ at that date. Crisis bands shade
the GFC (2008-Q3 through 2009-Q4) and COVID (March--December 2020).}
\label{fig:cumulative_loss}
\end{figure}

Third, the cumulative loss-differential plots reported in
Figure~\ref{fig:cumulative_loss} reveal that the forecasting gains
accumulate steadily over time rather than arising from a small number of
isolated events. The largest improvements occur during the global financial
crisis and COVID, precisely when uncertainty-duration effects should become
economically most important.

The comparison between linear and machine-learning estimators is also highly
revealing. The shrinkage estimators, Lasso, Ridge, and Elastic Net, 
track the performance of the full structural model relatively closely at
short and medium horizons ($+4\%$ to $+7\%$). However, these estimators
underperform at the monthly horizon because regularization shrinks the
cross-sectional and interaction features more aggressively than appears
economically appropriate.

The tree-based estimators perform considerably worse. Random Forest delivers
only weak improvements at short horizons and negative performance at the
monthly horizon. Gradient Boosting performs especially poorly, producing
forecasting deterioration between approximately $-10\%$ and $-19\%$.
These results carry an important methodological implication. The forecasting
gains arise primarily from economically structured feature construction
rather than from estimator flexibility itself. The likely explanation is
overfitting: the tree-based estimators operate in a relatively small
effective sample with approximately 18 predictors and substantial temporal
dependence. Under such conditions, flexible nonlinear algorithms may fit
noise and local idiosyncrasies rather than stable economic relationships.

\begin{figure}[H]
\centering
\includegraphics[width=\textwidth]{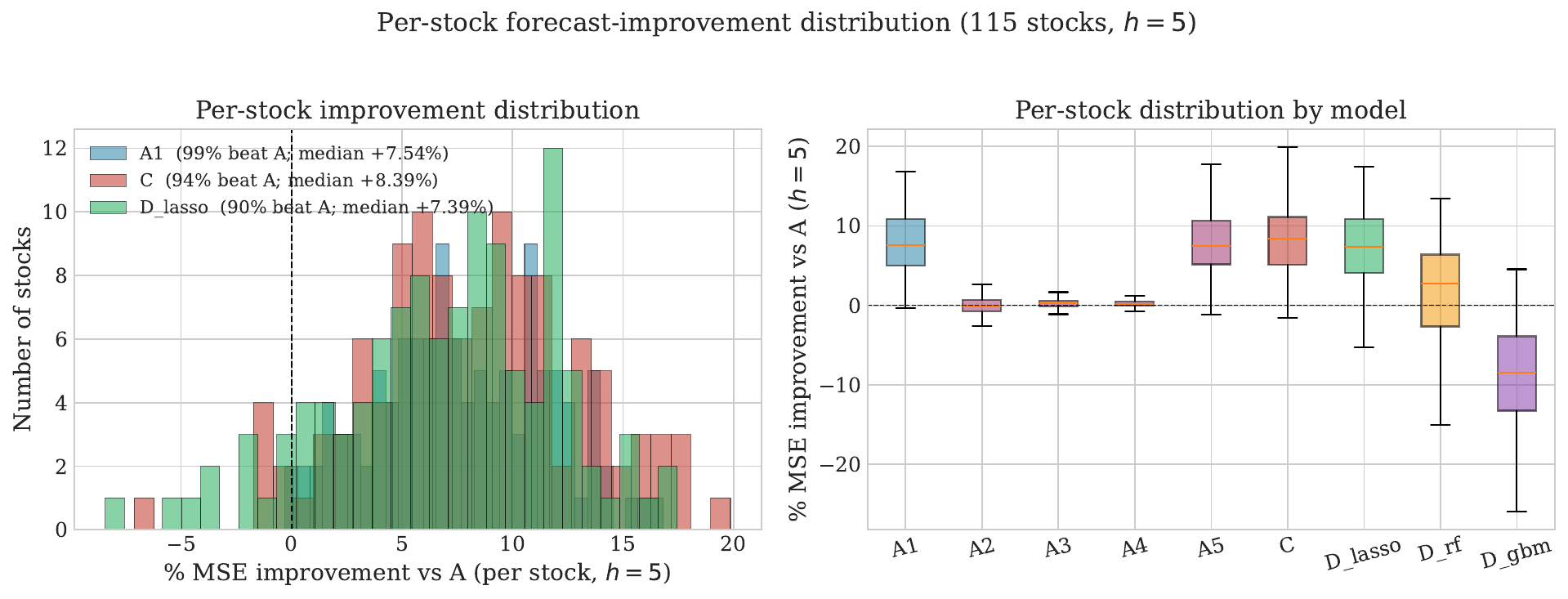}
\caption{Per-stock MSE improvement vs Model~A at $h=5$. Left panel:
histograms for $A_1$, $C$, and D-Lasso. Right panel: boxplot of the
per-stock improvement distribution across the full ladder.}
\label{fig:stock_distribution}
\end{figure}

Figure~\ref{fig:stock_distribution} further illustrates the breadth of the
forecasting gains across the cross-section of stocks. At the weekly
horizon, 99\% of stocks are improved by HAR-X relative to the HAR
benchmark, while approximately 94\% are improved by the full structural
model. By contrast, Gradient Boosting outperforms HAR for only a very
small fraction of the cross-section. This result demonstrates that the
persistence-based forecasting gains are not driven by a few isolated
securities. Rather, the gains appear broadly throughout the cross-section
of equities.

The layered forecasting results therefore support several central
conclusions. First, persistence behaves like an economically meaningful
state variable rather than merely a statistical nuisance parameter. Second,
persistence contributes most strongly during long-horizon and stress-regime
forecasting problems, precisely where uncertainty-duration effects should
matter most. Third, economically interpretable persistence structures appear
more important than highly flexible nonlinear estimation procedures.

\subsection{Feature Importance}

The forecasting results establish that persistence-related variables improve
predictive performance, particularly during stress regimes and at longer
horizons. The next question concerns which components of the persistence
feature vector actually drive these gains. To address this issue, the paper
examines standardized Lasso coefficients estimated on the full pooled
out-of-sample panel.

Table~\ref{tab:importance} reports the resulting coefficient estimates
separately for the daily, weekly, and monthly forecasting horizons. Because
all predictors are standardized prior to estimation, the coefficients are
directly comparable in magnitude.

Several important patterns emerge immediately.

First, the dominant predictors remain the classical HAR components:
$\log\RV^{(d)}$, $\log\RV^{(w)}$, $\log\RV^{(m)}$. Among these variables,
the monthly HAR component consistently carries the largest coefficient
magnitude, particularly at longer forecasting horizons. This result is
economically intuitive because long-horizon realized-volatility information
should matter most for medium- and long-horizon volatility forecasting.
The persistence of the HAR structure throughout all specifications
reinforces an important point emphasized earlier: traditional multi-horizon
volatility dynamics remain central even when persistence and roughness
variables are introduced. The persistence framework therefore complements
rather than replaces conventional HAR-type forecasting structures.

Second, the market-state variables, especially the VIX, also carry
a substantial coefficient magnitude. The standardized VIX coefficient remains
strongly positive across all horizons, confirming that broad
implied-volatility conditions provide important predictive information
regarding future realized volatility.

The third and most important finding concerns the persistence aggregates
themselves. Cross-sectional persistence measures such as $\overline d_t$
and $\sigma_d^t$ enter consistently across horizons with economically
meaningful coefficient magnitudes. Sectoral persistence averages
$\overline d_{s(i),t}$ also remain significant contributors to the
forecasting system. This result is crucial because it demonstrates that
persistence becomes economically informative primarily when viewed as a
system-wide or sector-wide state variable rather than as an isolated
individual-stock coefficient.

By contrast, the own-stock persistence estimate \(\widehat d_{i,t}\) is not eliminated by the 
Lasso procedure, but its role should be interpreted cautiously. The layered forecasting results 
show that own-stock persistence alone contributes little incremental forecasting gain once the 
HAR components are already included. Thus, the main economic contribution of the persistence 
block does not come from \(\widehat d_{i,t}\) in isolation, but from the way persistence is combined 
with market-wide stress, cross-sectional aggregation, sectoral persistence states, and 
persistence-by-stress interactions.

The persistence-by-stress interaction variable $\dhat_t\!\cdot\!\VIX_t$
also enters prominently throughout the forecasting system. Its coefficient
remains economically large across horizons, confirming that the forecasting
impact of market stress depends critically on the persistence state of the
financial system. This interaction structure is one of the central empirical
findings of the paper. High volatility by itself does not necessarily imply
prolonged uncertainty. However, high volatility occurring simultaneously
with elevated persistence indicates that uncertainty is likely to dissipate
slowly rather than quickly reverting to normal conditions.

The persistence-by-MOVE interaction also contributes positively, although
somewhat less strongly than the VIX interaction. This result suggests that
persistence interacts not only with equity-market stress but also with
broader fixed-income uncertainty conditions.

Several additional patterns deserve mention. The volatility-of-persistence
measure $\mathrm{Vol}(\dhat_t)$ enters negatively at longer horizons.
Economically, this may indicate that highly unstable persistence
environments reduce the predictive reliability of persistence itself. The
trend component $\mathrm{Trend}(\dhat_t)$ generally receives relatively
small coefficients, suggesting that the level of persistence matters more
than its short-run directional movement. The roughness variables $H_t$ and
$\Delta H_t$ also contribute more modestly than the persistence aggregates,
clarifying the relative roles of roughness and persistence. Roughness
captures local short-horizon irregularity, whereas persistence aggregates
appear to contain more economically relevant information regarding
medium-horizon uncertainty propagation.

Taken together, the feature-importance analysis provides a coherent
economic narrative. The dominant predictive structure remains the
traditional HAR multi-horizon volatility decomposition. Market-state
variables such as VIX provide substantial additional information regarding
current stress intensity. Persistence becomes economically important
primarily through its aggregate system-wide dimension and through its
interaction with market stress conditions. This result strongly supports
the structural interpretation developed throughout the paper. Persistence
functions as a state-dependent measure of how broadly and how persistently
uncertainty is propagating throughout the financial system. The feature
importance analysis also helps explain why the full structural model
outperforms the more flexible tree-based procedures: the forecasting gains
arise from a carefully structured economic feature space rather than from
highly nonlinear algorithmic complexity.


\begin{table}[htbp]
\centering
\caption{Standardised Lasso Coefficients on the Full Pooled Sample}
\label{tab:importance}
\small
\begin{tabular}{lccc}
\toprule
Feature & $h=1$ & $h=5$ & $h=22$ \\
\midrule
har\_d\_log & +0.1330 & +0.0991 & +0.0675 \\
har\_w\_log & +0.2106 & +0.1840 & +0.1460 \\
har\_m\_log & +0.3540 & +0.4047 & +0.4499 \\
ret\_lag1 & -0.0387 & -0.0474 & -0.0424 \\
ret\_lag1\_abs & +0.0241 & +0.0153 & +0.0177 \\
d\_gph & +0.0879 & +0.0837 & +0.0527 \\
delta\_d\_gph & +0.0113 & +0.0069 & +0.0029 \\
vol\_d\_gph & -0.0253 & -0.0394 & -0.0465 \\
trend\_d\_gph & -0.0076 & -0.0076 & -0.0000 \\
h & -0.0169 & -0.0091 & -0.0060 \\
delta\_h & +0.0043 & +0.0153 & +0.0094 \\
cs\_mean\_d & -0.0672 & -0.0474 & -0.0148 \\
cs\_std\_d & -0.0067 & -0.0058 & -0.0222 \\
sector\_mean\_d & +0.0147 & +0.0085 & +0.0095 \\
vix & +0.2784 & +0.2247 & +0.1132 \\
move & -0.0315 & +0.0000 & +0.0214 \\
d\_x\_vix & -0.1453 & -0.1432 & -0.1151 \\
d\_x\_move & +0.0227 & +0.0239 & +0.0381 \\
\bottomrule
\end{tabular}
\begin{tablenotes}\small
\item Notes: Pooled-Lasso fit on standardised features (full sample, all stocks, all out-of-sample dates), one fit per horizon. Coefficients are on standardised inputs and so are directly comparable in magnitude. Zero coefficients indicate features dropped by Lasso.
\end{tablenotes}
\end{table}

\subsection{Regime Split}
\label{subsec:regimes}

The structural interpretation developed earlier implies that persistence
measures should be most valuable precisely during periods in which
uncertainty resolves slowly. To evaluate this prediction directly, the
forecasting results are decomposed across several distinct market regimes:
low-volatility states, high-volatility states, and the COVID crisis period.

Table~\ref{tab:regime_split} reports the percentage mean squared error
improvement of each forecasting model relative to the HAR baseline within
these regimes. The results are striking and strongly support the economic
interpretation of persistence.

During low-VIX periods, the full structural model produces relatively
modest improvements over the HAR benchmark: $+2.0\%$, $+6.6\%$, $+9.7\%$
for the daily, weekly, and monthly horizons respectively. By contrast,
during high-VIX periods the improvements increase dramatically:
$+11.8\%$, $+17.9\%$, $+8.8\%$. The COVID period produces similarly
large gains: $+6.8\%$, $+18.9\%$, $+17.5\%$. These differences are
economically very large. The forecasting gains during crisis states are
often roughly three times larger than those observed during calm
volatility regimes.

Equally important is the decomposition of these gains across the layered
forecasting ladder. At the weekly horizon during high-VIX states, HAR-X
alone already produces substantial gains of $+16.1\%$. However, the full
structural model extends this improvement to approximately $+17.9\%$. The
difference, approximately 1.9 additional percentage points,
represents the incremental contribution of the persistence aggregates and
persistence-by-stress interaction terms. A similar pattern emerges during
COVID at the monthly horizon. HAR-X produces approximately $+15.8\%$
improvement, while the full structural model increases this to
approximately $+17.5\%$. Again, persistence contributes meaningful
incremental forecasting information beyond standard market-state variables
alone.

These results are exactly what the structural interpretation predicts. If
the cross-sectional persistence state $\overline d_t$ measures the duration
of unresolved uncertainty in the financial system, then persistence-related
variables should matter most precisely when uncertainty is elevated and slow
to dissipate. The gains are not concentrated within a few isolated event
windows. Rather, the improvements persist systematically throughout
prolonged stress periods, strongly suggesting that the forecasting gains
are capturing genuine economic dynamics rather than temporary statistical
artifacts.

\begin{figure}[H]
\centering
\includegraphics[width=\textwidth]{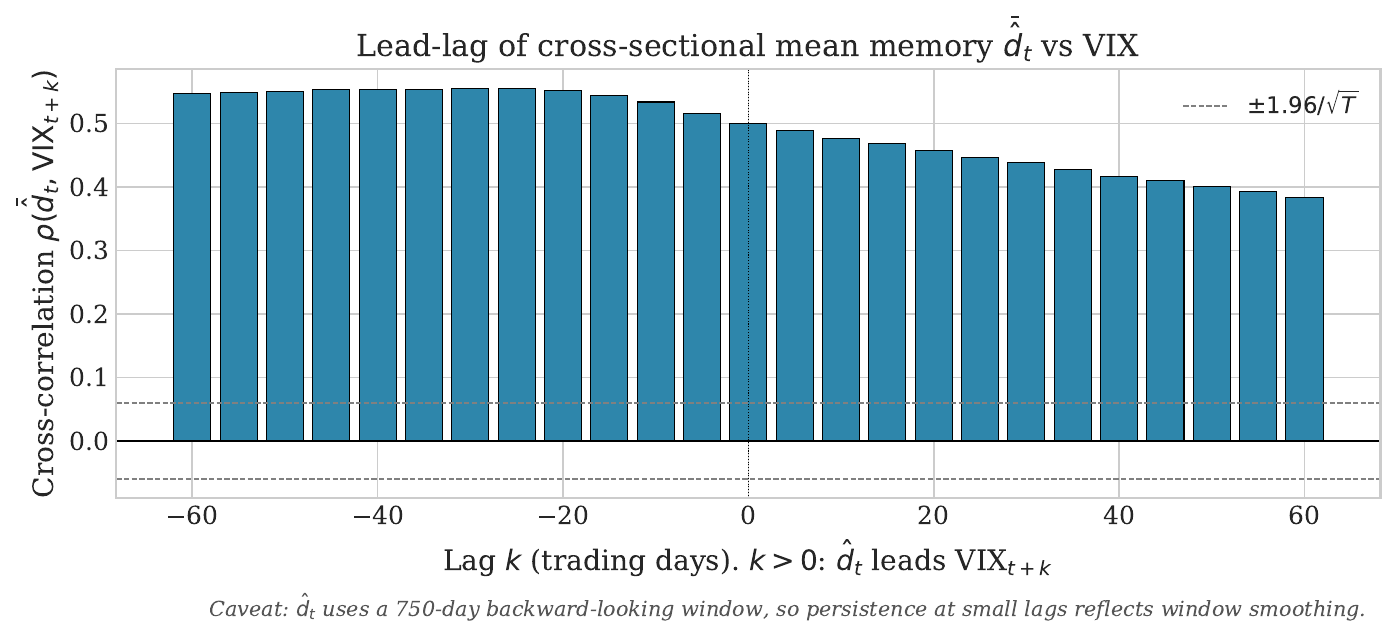}
\caption{Lead-lag cross-correlation between cross-sectional mean memory
$\bar{\hat d}_t$ and the VIX over $\pm 60$ trading days.}
\label{fig:leadlag}
\end{figure}

Figure~\ref{fig:leadlag} provides an additional diagnostic by examining
lead--lag correlations between the cross-sectional persistence state and
the VIX. The correlation function peaks near zero lag and remains
approximately symmetric around the contemporaneous point. This result
indicates that persistence and market stress co-move primarily
contemporaneously rather than through strong lead--lag relationships. The
absence of strong directional asymmetry suggests that persistence is
functioning as a coincident uncertainty-duration indicator rather than a
simple leading indicator of market volatility.

\begin{table}[htbp]
\centering
\caption{Regime-Conditioned MSE Improvement (\%) vs Model A}
\label{tab:regime_split}
\small
\resizebox{\textwidth}{!}{%
\begin{tabular}{llcccccccccccc}
\toprule
Regime & $h$ & A & A1 & A2 & A3 & A4 & A5 & C & D\_lasso & D\_ridge & D\_en & D\_rf & D\_gbm \\
\midrule
COVID (2020) & 1 & +0.00\% & +11.81\% & -1.52\% & -0.36\% & -1.41\% & +9.12\% & +6.83\% & +6.84\% & +4.56\% & +7.25\% & -8.04\% & -10.59\% \\
 & 5 & +0.00\% & +20.67\% & +0.07\% & -0.51\% & -2.52\% & +20.19\% & +18.88\% & +16.43\% & +13.72\% & +16.14\% & -0.48\% & +5.41\% \\
 & 22 & +0.00\% & +15.81\% & -3.74\% & +0.05\% & -5.82\% & +14.49\% & +17.49\% & +0.52\% & +0.49\% & +0.70\% & -7.04\% & -5.74\% \\
High VIX (Q4) & 1 & +0.00\% & +11.45\% & -1.30\% & +0.32\% & -0.51\% & +11.23\% & +11.84\% & +9.72\% & +8.29\% & +9.56\% & +3.53\% & -0.52\% \\
 & 5 & +0.00\% & +16.05\% & -0.66\% & +0.90\% & -0.49\% & +16.56\% & +17.94\% & +15.01\% & +13.51\% & +14.90\% & +6.31\% & +5.56\% \\
 & 22 & +0.00\% & +6.87\% & -1.10\% & +1.03\% & -1.47\% & +6.36\% & +8.75\% & +0.98\% & +0.68\% & +1.29\% & -7.62\% & -16.68\% \\
Low VIX (Q1) & 1 & +0.00\% & +3.03\% & -0.20\% & +0.14\% & +0.41\% & +3.07\% & +2.03\% & +1.53\% & +1.16\% & +1.53\% & -2.85\% & -21.87\% \\
 & 5 & +0.00\% & +5.58\% & +0.88\% & +1.75\% & +1.24\% & +5.75\% & +6.57\% & +5.29\% & +4.95\% & +5.35\% & -1.31\% & -17.32\% \\
 & 22 & +0.00\% & +5.24\% & +0.78\% & +2.94\% & +1.73\% & +5.18\% & +9.71\% & +3.69\% & +3.90\% & +3.87\% & -6.16\% & -17.36\% \\
\bottomrule
\end{tabular}%
}
\begin{tablenotes}\small
\item Notes: Out-of-sample MSE improvement on $\log RV^{PK}$ relative to Model A, computed within each regime. VIX quartiles are computed on the out-of-sample evaluation window. Crisis windows are 2008-Q3 to 2009-Q4 (GFC) and Mar--Dec 2020 (COVID). Positive values indicate the model beats HAR within the regime.
\end{tablenotes}
\end{table}

The regime analysis therefore provides some of the strongest evidence
supporting the economic interpretation of persistence developed earlier in
the paper. Persistence-based forecasting variables contribute relatively
little during calm markets because uncertainty resolves quickly and the HAR
structure already captures most predictable volatility behavior. During
prolonged stress episodes, however, persistence aggregates become highly
informative because they summarize how slowly the financial system is
absorbing uncertainty. These findings strongly reinforce the interpretation
of persistence as a state-dependent measure of uncertainty duration rather
than merely a statistical parameter governing autocorrelation decay.

\subsection{Sector Split}

The regime analysis established that persistence features become most
valuable during broad stress episodes. The next question concerns whether
the forecasting gains are distributed uniformly across industries or
whether certain sectors benefit more strongly from persistence-based
information.

Table~\ref{tab:sector_split} reports the sector-by-sector forecasting
improvements for the weekly forecasting horizon $h=5$, measured relative
to the HAR benchmark. The results reveal substantial heterogeneity across
sectors, but every GICS sector exhibits positive gains under the full
structural model.

The strongest improvements occur in Materials, Industrials, Real Estate,
and Financials. The percentage mean squared error reductions for these
sectors equal approximately $+12.7\%$, $+10.5\%$, $+10.3\%$, $+9.5\%$.
By contrast, the smallest improvements occur in Information Technology and
Energy, with gains approximately equal to $+4.2\%$ and $+4.9\%$.
Although still positive, these improvements are materially smaller than
those observed in the balance-sheet-sensitive sectors.

These sectoral patterns are economically highly plausible. Financials and
Real Estate are naturally sensitive to systemic funding conditions, leverage
cycles, collateral constraints, and broad credit-market stress. Their
volatility dynamics therefore depend heavily on prolonged uncertainty
propagation throughout the financial system. Industrials and Materials
display similar behavior because they are highly exposed to macroeconomic
cycles, global demand conditions, and prolonged investment uncertainty.
Volatility shocks in these sectors often propagate slowly through
inventories, capital expenditures, commodity markets, and financing
conditions.

By contrast, Information Technology volatility is often driven more
strongly by firm-specific innovation shocks, product announcements, and
idiosyncratic news flow rather than by prolonged systemic stress
propagation. Persistence aggregates therefore contribute less incremental
forecasting information once conventional volatility predictors are already
included. The Energy sector provides another informative case. Energy
volatility is heavily influenced by oil-price dynamics and
commodity-specific macroeconomic factors that are not explicitly modeled
within the forecasting framework. Persistence-based financial-system
variables therefore explain a smaller share of volatility behavior in this
sector.

\begin{figure}[H]
\centering
\includegraphics[width=\textwidth]{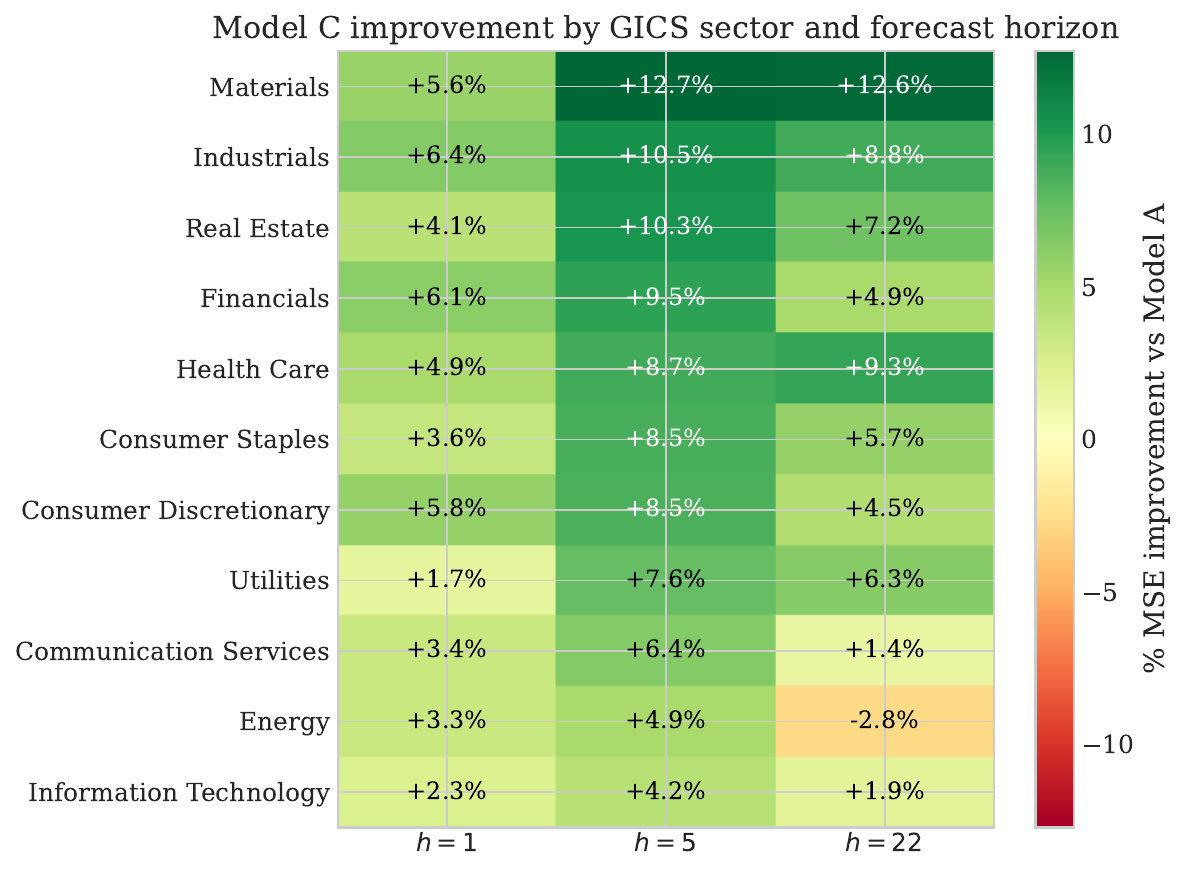}
\caption{Percentage MSE improvement of Model~$C$ over Model~$A$, by
GICS sector and forecast horizon. Sectors are sorted by the $h=5$ gain.}
\label{fig:sector_heatmap}
\end{figure}

Figure~\ref{fig:sector_heatmap} extends the sectoral comparison across all
three forecasting horizons. First, the sectoral ranking remains remarkably
stable across horizons. Materials, Industrials, Real Estate, and Financials
remain among the strongest beneficiaries throughout the daily, weekly, and
monthly horizons.

Second, the gains generally become larger at the monthly horizon for the
balance-sheet-sensitive sectors. For example, Materials displays
approximately $+12.6\%$ improvement at $h=22$, while Industrials and Health
Care also exhibit particularly strong long-horizon gains. This pattern
especially reinforces the structural interpretation: persistence-based
variables appear most useful precisely in sectors where uncertainty
propagation is likely to persist over extended horizons.

Third, several sectors display asymmetric horizon behavior. Utilities, for
example, show relatively modest short-horizon improvements but stronger
medium- and long-horizon gains. This may reflect the gradual adjustment
dynamics associated with regulated industries and slower-moving
macroeconomic factors.
The sector analysis also provides additional evidence against purely
statistical interpretations of the forecasting gains. If the persistence
variables were merely overfitting generic volatility dynamics, one would
expect relatively uniform gains across sectors. Instead, the gains align
closely with economically plausible sectoral exposure to systemic stress
propagation, leverage conditions, and uncertainty duration.

The layered forecasting ladder further clarifies the source of these gains.
In nearly every sector, HAR-X alone captures a substantial portion of the
overall improvement. However, the full structural model consistently extends
the gains beyond HAR-X, especially within the sectors most exposed to
prolonged systemic uncertainty. The shrinkage estimators generally track
the sectoral ranking reasonably well, although their gains are somewhat
attenuated. The tree-based estimators again perform poorly across nearly
all sectors, with Gradient Boosting frequently generating strongly negative
forecasting performance. These results reinforce one of the central
methodological conclusions of the paper: the forecasting gains arise
primarily from economically interpretable persistence structures rather than
from flexible nonlinear estimation procedures.

\begin{table}[htbp]
\centering
\caption{Sector-Level MSE Improvement (\%) vs Model A at $h=5$}
\label{tab:sector_split}
\small
\resizebox{\textwidth}{!}{%
\begin{tabular}{lcccccccccc}
\toprule
Sector & A & A1 & A2 & A3 & A4 & A5 & C & D\_lasso & D\_rf & D\_gbm \\
\midrule
Communication Services & +0.00\% & +7.29\% & -0.76\% & +0.42\% & +0.38\% & +7.37\% & +6.39\% & +5.76\% & +0.53\% & -14.61\% \\
Consumer Discretionary & +0.00\% & +8.36\% & -0.03\% & +0.11\% & +0.26\% & +7.44\% & +8.51\% & +6.97\% & +1.66\% & -9.57\% \\
Consumer Staples & +0.00\% & +7.22\% & +0.02\% & +0.51\% & +0.33\% & +6.83\% & +8.55\% & +6.46\% & +0.86\% & -10.85\% \\
Energy & +0.00\% & +4.28\% & +0.71\% & +0.13\% & +0.14\% & +5.20\% & +4.91\% & +1.73\% & -2.84\% & -13.14\% \\
Financials & +0.00\% & +9.45\% & +0.27\% & +0.19\% & +0.19\% & +9.46\% & +9.48\% & +7.79\% & +4.24\% & -6.76\% \\
Health Care & +0.00\% & +6.62\% & -0.60\% & +0.61\% & +0.11\% & +7.12\% & +8.75\% & +7.87\% & +3.20\% & -9.00\% \\
Industrials & +0.00\% & +9.10\% & -0.29\% & +0.79\% & +0.25\% & +9.41\% & +10.51\% & +8.30\% & +3.39\% & -8.15\% \\
Information Technology & +0.00\% & +5.07\% & -0.26\% & +0.15\% & +0.26\% & +4.55\% & +4.19\% & +3.51\% & -0.61\% & -14.55\% \\
Materials & +0.00\% & +10.99\% & -0.25\% & +0.61\% & +0.01\% & +10.89\% & +12.72\% & +11.28\% & +4.10\% & -12.16\% \\
Real Estate & +0.00\% & +10.22\% & +0.79\% & -0.04\% & -0.09\% & +10.16\% & +10.29\% & +10.50\% & +5.66\% & -7.99\% \\
Utilities & +0.00\% & +8.08\% & +1.45\% & +0.23\% & +0.22\% & +6.98\% & +7.62\% & +9.15\% & +2.58\% & -4.60\% \\
\bottomrule
\end{tabular}%
}
\begin{tablenotes}\small
\item Notes: Out-of-sample MSE improvement on $\log RV^{PK}$ at horizon $h=5$, by GICS sector, relative to Model A. Pooled across stocks within each sector and over the entire out-of-sample window.
\end{tablenotes}
\end{table}

\subsection{Robustness}

The final step in the empirical analysis is to determine whether the
forecasting improvements remain stable across alternative estimation
procedures, rolling-window choices, volatility proxies, liquidity
conditions, and benchmark specifications.

Table~\ref{tab:robustness} summarizes a broad robustness battery for the
headline comparison between the full structural model and the HAR benchmark
at the weekly forecasting horizon. The results are remarkably stable across
all perturbations considered.

\begin{table}[htbp]
\centering
\caption{Robustness Summary: Model C vs Model A, $h=5$}
\label{tab:robustness}
\small
\resizebox{\textwidth}{!}{%
\begin{tabular}{lccc}
\toprule
Variant & MSE & \%$\Delta$ vs A & HLN DM-$t$ \\
\midrule
Headline (Model C, full panel) & 0.3334 & +8.24\% & $+3.87$ \\
Inference: plain DM (illustrative) & 0.3334 & +8.24\% & $+3.87$ (plain $+23.97$) \\
Benchmark: GARCH(1,1) on returns & 0.7033 & -93.54\% & $-20.30$ \\
Liquidity: low-liquidity half & 0.3342 & +8.95\% & $+4.26$ \\
Liquidity: high-liquidity half & 0.3327 & +7.51\% & $+3.33$ \\
Estimator: $\hat d_{LW}$ replaces $\hat d_{GPH}$ & 0.3336 & +8.21\% & $+3.85$ \\
Rolling window: 500 days & 0.3302 & +9.13\% & $+4.00$ \\
Rolling window: 1000 days & 0.3319 & +8.67\% & $+3.63$ \\
Target: log mean future squared returns & 0.9102 & +4.16\% & $+3.10$ \\
Regime: high VIX (Q4) & -- & +17.94\% & -- \\
Regime: low VIX (Q1) & -- & +6.57\% & -- \\
Regime: COVID 2020 & -- & +18.88\% & -- \\
\bottomrule
\end{tabular}%
}
\begin{tablenotes}\small
\item Notes: Robustness checks for Model C against Model A at $h=5$ on $\log RV^{PK}$. Liquidity halves are formed from the static median of mean inverse dollar volume across the sample. $\hat d_{LW}$ replaces $\hat d_{GPH}$ in the entire feature block (including memory dynamics and interactions). Window variants refit the rolling LRD estimation at the alternative window before regenerating the full feature stack. The target variant replaces $\log RV^{PK}$ with log mean future squared returns (Models A and C both refit). The GARCH(1,1) row fits a constant-mean GARCH(1,1) on log returns with refit every 20 sample steps and reports the resulting log mean conditional variance forecast against the same Parkinson target; its forecast scale differs from the Parkinson RV target (returns variance vs.\ range-based variance proxy), so the negative $\%\Delta$ vs A reflects both this level mismatch and the well-known limitation of returns-only GARCH for forecasting realised variance. Under the proxy-robust QLIKE loss of \citet{patton2011volatility} the GARCH(1,1) HLN-corrected DM statistic against Model~A narrows to $-2.80$ ($p=0.005$), versus $-20.30$ under MSE in the column above. Regime rows come from the Table~\ref{tab:regime_split} split at $h=5$. The plain-DM stat is the unscaled iid pooled-cell statistic; the HLN stat is the panel-aware Harvey-Leybourne-Newbold finite-sample-corrected version.
\end{tablenotes}
\end{table}
Replacing the GPH persistence estimator throughout the feature set with
the local-Whittle estimator produces forecasting improvements almost
identical to the headline specification: $+8.21\%$ versus the baseline
$+8.24\%$. This result is important because it demonstrates that the
forecasting gains are not driven by the particular semiparametric estimator
used to construct the persistence variables.

The rolling-window robustness checks produce similarly stable results.
Using a shorter 500-day rolling estimation window yields approximately
$+9.13\%$ forecast improvement, while using a longer 1000-day window
yields approximately $+8.67\%$. Both estimates remain highly statistically
significant. These findings indicate that the forecasting gains are not
artifacts of a particular rolling-window length. Persistence-based
forecasting variables remain informative across substantially different
temporal smoothing structures.

The volatility-proxy robustness checks are also revealing. Replacing the
Parkinson realized-variance target with the logarithm of future mean
squared returns, a substantially noisier volatility proxy, reduces
the magnitude of the forecasting gains but still produces statistically
significant improvement: $+4.16\%$. This result suggests that the
persistence-based forecasting gains are not specific to the Parkinson
volatility estimator itself. However, the smaller improvement also confirms
that cleaner volatility measurement substantially improves the ability to
extract economically meaningful persistence information.

Liquidity-based sample splits produce additional evidence supporting the
structural interpretation. Dividing the sample into high- and
low-liquidity halves yields improvements approximately equal to $+8.95\%$
and $+7.51\%$. The somewhat larger gains in the low-liquidity sample are
economically plausible because volatility persistence should become more
important when markets absorb shocks more slowly and liquidity provision
is weaker.

The GARCH(1,1) benchmark row warrants comment. A conventional GARCH(1,1)
model estimated directly on returns performs substantially worse than the
HAR benchmark when evaluated against the Parkinson realized-volatility
target: $-93.5\%$ relative performance under mean squared error loss.
However, under the proxy-robust QLIKE loss function, which is more
robust when volatility proxies contain measurement noise, the relative
deterioration becomes much smaller (HLN DM-$t = -2.80$). This distinction
is important because it highlights a broader econometric issue. Direct
comparisons between returns-based conditional variance models and
range-based realized-volatility proxies involve substantial scale mismatch
and measurement differences. The paper therefore emphasizes that the
forecasting comparison should primarily be interpreted within the
realized-volatility forecasting environment rather than as a definitive
rejection of all conditional-variance models.

Finally, the robustness analysis again illustrates the importance of
correct statistical inference. The naive pooled-cell Diebold--Mariano
statistic ignoring cross-sectional dependence equals approximately
$+23.97$, whereas the properly panel-aware HLN-corrected statistic equals
approximately $+3.87$. This difference is enormous and demonstrates how
severely significance levels may be inflated when within-date
cross-sectional dependence is ignored.

Taken together, the robustness analysis strongly supports the central
empirical findings of the paper. The persistence-based forecasting gains
remain stable across alternative persistence estimators, rolling-window
lengths, volatility proxies, liquidity conditions, and evaluation
procedures. The gains are therefore unlikely to be artifacts of a
particular econometric specification or isolated sample feature. Most
importantly, the robustness results continue to support the broader
structural interpretation developed throughout the paper. Persistence
measures behave coherently across market states, forecasting environments,
and estimation procedures, suggesting that they capture economically
meaningful features of uncertainty propagation rather than merely
statistical regularities.

\section{Economic Interpretation and Implications}
\label{sec:economic}

The empirical results of Section~\ref{sec:results} are important not
simply because they improve a forecasting statistic, but because they
clarify the economic meaning of persistence itself. If persistence-based
variables systematically improve volatility forecasts across crisis periods,
high-stress states, and balance-sheet-sensitive sectors, then persistence
should not be interpreted merely as a technical characteristic of a
stochastic process. Instead, it should be understood as an empirical
indicator of how slowly financial markets absorb and resolve uncertainty.

The central interpretation developed throughout the paper is therefore
economic rather than purely statistical. The rolling persistence state
summarizes the expected duration of volatility shocks and uncertainty
propagation throughout the financial system. Under this interpretation,
persistence measures are not simply filters embedded within econometric
specifications. They are state variables describing the temporal
organization of market stress.

\subsection{Volatility-Managed Portfolios}
\label{subsec:volmanaged}

The first direct economic application concerns volatility-managed
portfolios. Following \citet{moreira2017volatility}, the paper constructs
volatility-managed equity portfolios that scale exposure inversely with
forecast variance. For stock $i$ at date $t$, the managed five-day
return is defined as
\begin{equation}
  f_{m,i,t} \;=\; \frac{c_{m,i}}{\hat\sigma^2_{m,i,t}}\,r_{i,t\to t+5},
\end{equation}
where $\hat\sigma^2_{m,i,t} = \exp(\hat y_{m,i,t})$ denotes the
model-based forecast variance and $c_{m,i}$ rescales the managed return to
preserve unconditional variance comparability with buy-and-hold strategies.

The resulting portfolio evidence is economically highly informative. Over
the full sample, the unmanaged equal-weight buy-and-hold portfolio produces
a Sharpe ratio approximately equal to $0.73$, while the
persistence-augmented managed strategy also produces approximately $0.73$.
At first glance, this result may appear disappointing. However, the
full-sample average conceals the central economic role of persistence. The
benefits emerge primarily during stress states rather than during calm
market conditions.

During COVID, the unmanaged portfolio Sharpe ratio collapses to
approximately $0.65$, while the persistence-augmented strategy produces a
Sharpe ratio of $1.36$. Similarly, the certainty-equivalent return for a
mean-variance investor with risk aversion parameter $\gamma = 5$ moves from
approximately $-3.0\%$ under buy-and-hold to approximately $+12.3\%$ under
the persistence-augmented strategy. These improvements are economically
very large.

The layered forecasting structure again helps isolate the source of the
gains. In COVID, for example: Sharpe$(A) = 0.91$, Sharpe$(A_1) = 1.06$,
Sharpe$(C) = 1.36$. The persistence aggregates therefore generate an
incremental Sharpe improvement of roughly $0.30$ even after accounting for
the market-state effects captured by HAR-X.

\begin{figure}[H]
\centering
\includegraphics[width=\textwidth]{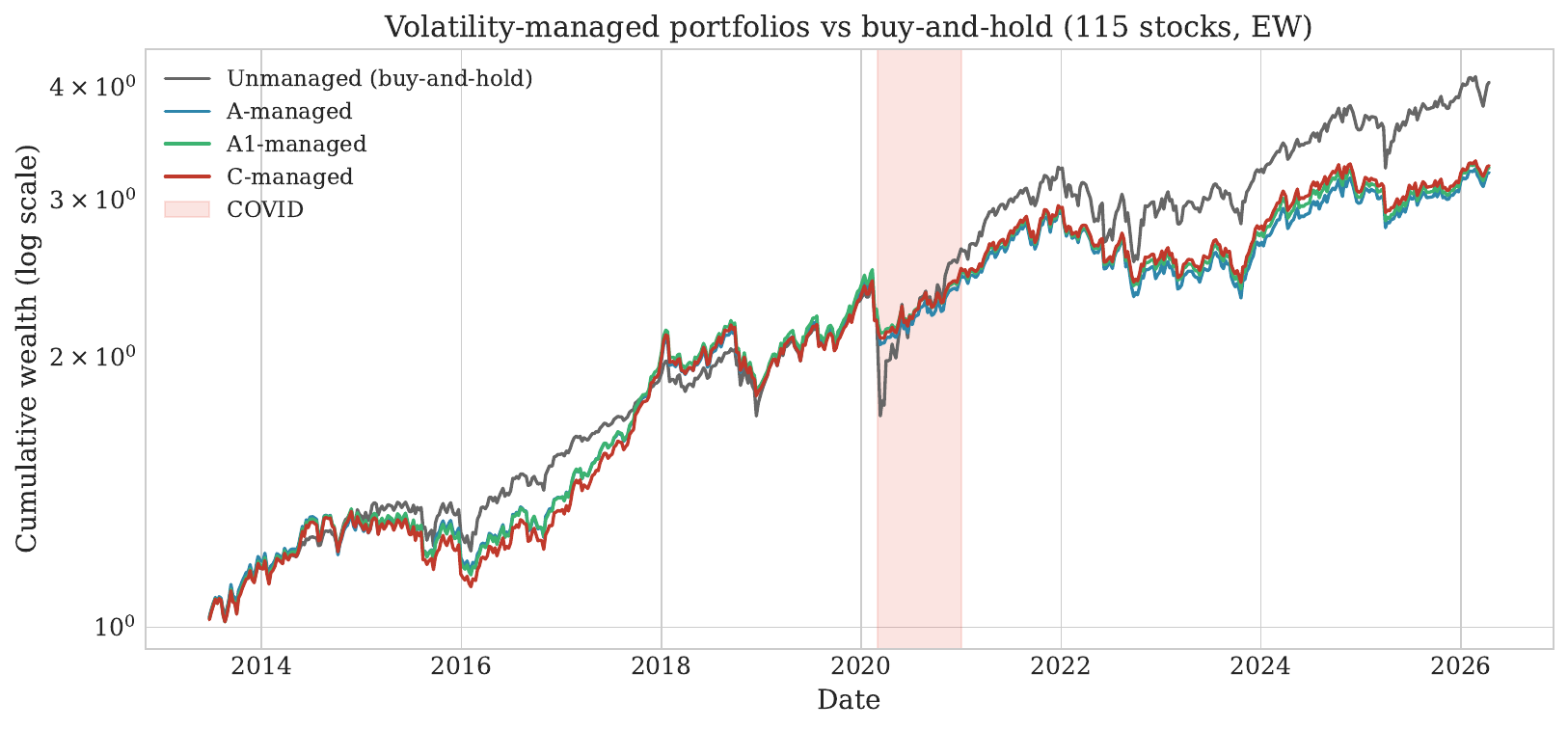}
\caption{Cumulative wealth (log scale) of the unmanaged equal-weight
buy-and-hold portfolio and three volatility-managed variants on the
out-of-sample window. Shaded band marks COVID (March--December 2020).}
\label{fig:volmanaged}
\end{figure}

Figure~\ref{fig:volmanaged} provides a particularly clear visual
interpretation of the mechanism. The managed portfolios sharply reduce
exposure precisely during periods in which forecast variance spikes upward.
The persistence-augmented strategy cuts exposure most aggressively during
the global financial crisis and COVID, preserving capital during the
collapse and allowing more efficient participation during the subsequent
recovery. This evidence strongly supports the economic interpretation of
persistence as a duration-of-stress indicator. Traditional volatility
forecasts capture the intensity of contemporaneous risk. Persistence-based
variables additionally capture how long elevated uncertainty conditions are
likely to remain active.


\begin{table}[htbp]
\centering
\caption{Volatility-Managed Portfolios: Risk-Adjusted Performance (Moreira-Muir 2017 Construction)}
\label{tab:volmanaged}
\small
\begin{tabular}{llccccc}
\toprule
Regime & Portfolio & Ann.\ Ret. & Ann.\ Vol. & Sharpe & Max DD & CER \\
\midrule
Full sample & Unmanaged (buy-and-hold) & +10.94\% & +14.89\% & +0.73 & -28.45\% & +5.40\% \\
 & A-managed & +9.13\% & +12.91\% & +0.71 & -19.89\% & +4.96\% \\
 & A1-managed & +9.22\% & +12.84\% & +0.72 & -18.56\% & +5.10\% \\
 & C-managed & +9.27\% & +12.72\% & +0.73 & -17.92\% & +5.23\% \\
\midrule
Low VIX (Q1) & Unmanaged (buy-and-hold) & +6.71\% & +7.69\% & +0.87 & -13.19\% & +5.23\% \\
 & A-managed & +9.29\% & +11.65\% & +0.80 & -20.54\% & +5.90\% \\
 & A1-managed & +10.65\% & +12.12\% & +0.88 & -20.18\% & +6.98\% \\
 & C-managed & +11.37\% & +12.41\% & +0.92 & -20.21\% & +7.52\% \\
\midrule
High VIX (Q4) & Unmanaged (buy-and-hold) & +21.02\% & +23.19\% & +0.91 & -24.12\% & +7.58\% \\
 & A-managed & +11.63\% & +13.55\% & +0.86 & -16.71\% & +7.04\% \\
 & A1-managed & +10.87\% & +12.39\% & +0.88 & -14.68\% & +7.04\% \\
 & C-managed & +11.23\% & +12.08\% & +0.93 & -14.35\% & +7.58\% \\
\midrule
COVID 2020 & Unmanaged (buy-and-hold) & +19.62\% & +30.07\% & +0.65 & -11.96\% & -2.98\% \\
 & A-managed & +11.03\% & +12.09\% & +0.91 & -4.22\% & +7.38\% \\
 & A1-managed & +11.48\% & +10.82\% & +1.06 & -4.33\% & +8.55\% \\
 & C-managed & +15.54\% & +11.39\% & +1.36 & -5.04\% & +12.30\% \\
\midrule
\bottomrule
\end{tabular}
\begin{tablenotes}\small
\item Notes: Equal-weight portfolio of 115 stocks, weekly ($h=5$ trading days) rebalancing aligned to the forecast sample stride. The volatility-managed portfolio uses position weight $w_{m,i,t} = c_{m,i} / \hat\sigma^2_{m,i,t}$ where $\hat\sigma^2_{m,i,t} = \exp(\hat y_{m,i,t})$ is the model-$m$ forecast of mean $RV^{PK}$ over $[t+1, t+5]$. The constant $c_{m,i}$ is set so that $\mathrm{Var}(f_{m,i}) = \mathrm{Var}(r_i)$ at the stock level. Annualized assuming $252/5$ five-day periods per year. CER computed for a mean-variance investor with risk aversion $\gamma = 5$.
\end{tablenotes}
\end{table}

\subsection{Persistence as a Temporal Stress Indicator}

The empirical evidence suggests that persistence measures summarize a
fundamentally different dimension of financial stress than conventional
volatility indicators. Two periods may exhibit similar instantaneous
volatility levels while differing substantially in the expected duration of
uncertainty. One market may experience a brief volatility spike followed
by rapid stabilization. Another may experience prolonged uncertainty
propagation through funding markets, leverage channels, and institutional
balance sheets. Persistence measures appear to distinguish between these
environments.

The cross-sectional persistence state $\overline d_t$ captures this
temporal dimension directly. Empirically, the cross-sectional mean
persistence measure rises approximately 68\% during the global financial
crisis and approximately 86\% during COVID relative to the calm 2013--2014
baseline. Its contemporaneous correlation with the VIX equals approximately
$+0.50$. The persistence-by-stress interaction variable strengthens this
interpretation even further. For AAPL, for example, the interaction
$\dhat_t\!\cdot\!\VIX_t$ rises from approximately $1.05$ during calm
periods to approximately $12.16$ during the global financial crisis and
approximately $8.22$ during COVID. These magnitudes are economically
enormous and strongly inconsistent with the interpretation of persistence
as a purely technical econometric parameter. Instead, persistence appears
to summarize how deeply and how persistently uncertainty is propagating
throughout the financial system.

\subsection{Implications for Risk Management and Stress Testing}

The forecasting gains at longer horizons carry important implications for
risk management. Traditional volatility forecasting systems are central to
Value-at-Risk, expected shortfall, margin determination, stress testing,
capital allocation, and portfolio scaling. Most such systems focus
primarily on the expected magnitude of future volatility rather than on its
expected duration. The results of the paper suggest that this may be
incomplete.

At the monthly horizon ($h=22$), the persistence-augmented model improves
forecasting performance by approximately $+5.5\%$ overall and approximately
$+17.5\%$ during COVID. These gains indicate that persistence contains
forward-looking information about the duration of elevated volatility
conditions that is not captured by standard HAR or HAR-X structures. This
information is particularly relevant for horizon-sensitive risk-management
problems, including:
\begin{itemize}
  \item multi-day expected shortfall,
  \item margin calibration,
  \item stress-duration scenarios,
  \item liquidity horizon adjustments, and
  \item multi-period capital allocation.
\end{itemize}
Persistence therefore provides information not merely about how volatile
markets currently are, but about how long elevated risk conditions are
likely to persist.

\subsection{Portfolio Allocation and Volatility Trading}

Persistence also has important implications for portfolio management and
volatility trading. Volatility-managed strategies implicitly assume that
volatility forecasts can distinguish temporary volatility spikes from
persistent regime changes. Forecasting systems that fail to distinguish
these environments may reduce exposure after temporary shocks and re-enter
markets too early, producing costly regime-misclassification errors.
Persistence-based variables help mitigate this problem because they
summarize the expected duration of uncertainty conditions rather than
merely their current intensity.

This interpretation also carries implications for volatility trading across
maturities. When contemporaneous volatility is elevated but persistence
remains low, uncertainty may dissipate rapidly, favoring shorter-dated
volatility exposure. By contrast, a high-persistence state implies that
medium-horizon implied volatility may remain elevated for prolonged periods.
Persistence therefore provides a potential economic discipline for
interpreting volatility-term-structure dynamics.

\subsection{Macroprudential Monitoring}

The persistence framework also suggests broader macroprudential
applications. Most macroprudential monitoring systems focus on
contemporaneous indicators such as leverage, credit spreads, liquidity
conditions, and realized volatility. These indicators measure current
stress but do not directly summarize the expected duration of stress
propagation. Persistence measures potentially fill this gap.

A simultaneous increase in $\overline d_t$ and $\sigma_d^t$ may indicate
that uncertainty is both highly persistent and unevenly distributed
throughout the financial system. This combination is especially dangerous
because it suggests broad systemic vulnerability together with localized
pockets of acute fragility. Although the present paper does not explicitly
develop a macroprudential framework, the persistence-state variables
introduced here naturally extend to broader financial-stability
applications.

\subsection{Implications for Machine Learning in Finance}

Finally, the forecasting comparisons carry broader methodological
implications for machine learning in finance. The linear, economically
structured forecasting systems consistently outperform the highly flexible
tree-based estimators throughout the empirical analysis. Random Forest
contributes little incremental value, while Gradient Boosting substantially
worsens forecasting performance across nearly all specifications and
regimes.

This result strongly suggests that the economic value arises primarily from
feature-space construction rather than estimator complexity itself. The
persistence variables become useful because they summarize economically
interpretable mechanisms , uncertainty duration, stress propagation,
cross-sectional fragility, and persistence-by-stress interactions ,
rather than because nonlinear algorithms discover arbitrary predictive
patterns in large feature spaces. The results therefore support a broader
methodological perspective increasingly emphasized in empirical finance:
machine learning contributes most effectively when guided by economically
interpretable feature engineering rather than by unrestricted algorithmic
flexibility alone.

\section{Conclusion}
\label{sec:conclusion}

This paper has argued that persistence in financial volatility should not
be treated merely as a technical property of a stochastic process. Instead,
persistence should be interpreted as an economically meaningful feature of
financial markets reflecting how slowly uncertainty is absorbed, how long
the effects of shocks remain active, and how market conditions evolve
across calm and stressed regimes.

The classical econometric literature introduced persistence through
fractional integration, long-memory dependence, and conditional-volatility
models, providing rigorous tools for describing slow decay in volatility
dynamics. The broader contribution of the present paper has been to
reinterpret these measurements from a reduced-form perspective. The memory
parameter is not only a coefficient governing autocorrelation decay. It is
also a reduced-form indicator of information persistence, liquidity-mediated
propagation, heterogeneous adjustment across investors and sectors, and the
temporal depth of market stress.

A second central theme has been that volatility dynamics contain multiple
temporal layers. At short horizons, volatility exhibits highly irregular
local behavior emphasized by the rough-volatility literature
\citep{gatheral2018volatility}. At longer horizons, volatility displays
persistent dependence captured by long-memory and multi-horizon forecasting
structures such as HAR. These dimensions are not mutually exclusive. The
empirical analysis demonstrates that financial volatility is simultaneously
rough at short horizons and persistent over longer horizons.

The empirical evidence is remarkably strong. The 115-stock panel covering
November 2001 through April 2026 produces cross-sectional GPH and
local-Whittle persistence estimates of approximately
\begin{equation*}
  \dhat = 0.226 \quad \text{and} \quad \dhat = 0.440,
\end{equation*}
with statistical significance in nearly every stock. At the same time, the
rolling Hurst exponent satisfies $H < \tfrac12$ throughout the panel, with
a cross-sectional mean near $H = 0.063$. These findings confirm that
volatility exhibits both long-horizon persistence and substantial local
roughness simultaneously.

A third contribution has been operational and forecasting-oriented.
Embedding persistence-based variables into a layered forecasting framework
produces a clean empirical attribution of predictive gains. Own-stock
persistence by itself contributes little incremental forecasting information
because the HAR structure already captures much of the predictable
individual volatility persistence. However, cross-sectional persistence
aggregates, sectoral persistence states, and persistence-by-stress
interactions provide substantial incremental forecasting improvements beyond
both HAR and HAR-X benchmarks.

The forecasting improvements are economically and statistically meaningful:
4.6--8.2\% out-of-sample MSE reduction across horizons relative to HAR,
with significance at conventional levels under panel-aware HLN-corrected
Diebold--Mariano inference. The gains become particularly strong during
stress regimes: $+17.9\%$ during high-VIX states and $+18.9\%$ during
COVID at the weekly horizon.

The economic significance extends beyond forecasting statistics themselves.
In the Moreira--Muir volatility-managed portfolio exercise, the
persistence-augmented forecasting system produces a COVID Sharpe ratio of
$1.36$ compared with $0.65$ for buy-and-hold and $1.06$ for the
HAR-X-managed benchmark. The persistence aggregates therefore generate a
substantial incremental Sharpe improvement during stress states.

Equally important is the methodological implication concerning machine
learning in finance. The tree-based machine-learning procedures fail to
match the performance of the economically structured forecasting systems.
Random Forest contributes little incremental value, while Gradient Boosting
substantially worsens forecasting performance. The forecasting gains
therefore arise primarily from economically interpretable feature
construction rather than from unrestricted nonlinear estimator flexibility.
This finding supports a broader methodological lesson increasingly
emphasized in empirical finance: machine learning contributes most
effectively when guided by economically interpretable feature engineering
rather than by purely algorithmic complexity.

The paper also suggests several directions for future research. One natural
extension is to enrich the rough-volatility component using longer
high-frequency datasets in order to study more carefully how local
irregularity interacts with long-horizon persistence across market regimes.
Another extension would apply the framework beyond U.S.\ equities to fixed
income, foreign exchange, commodities, and volatility derivatives, where
the structural interpretation of persistence may differ but is likely to
remain economically informative.

A third direction would integrate the cross-sectional persistence state
into multivariate covariance and correlation forecasting systems. Because
many practical portfolio-allocation and risk-management problems involve
portfolios rather than individual assets, this extension would connect
naturally to the macroprudential literature on systemic fragility and
financial stability.

More broadly, the results suggest that persistence measures deserve
interpretation not merely as statistical objects but as empirical
indicators of the temporal organization of financial uncertainty. Markets
differ not only in how volatile they are, but also in how long uncertainty
remains active once shocks occur. Persistence-based state variables appear
capable of measuring this dimension directly.

\paragraph{Acknowledgements.}
We thank Texas Tech University for institutional support.

\bibliographystyle{apalike}
\bibliography{refs}

@article{andersen2003modeling,
  author  = {Andersen, Torben G. and Bollerslev, Tim and Diebold, Francis X. and Labys, Paul},
  title   = {Modeling and Forecasting Realized Volatility},
  journal = {Econometrica},
  volume  = {71},
  number  = {2},
  pages   = {579--625},
  year    = {2003},
}

@article{baillie1996fractionally,
  author  = {Baillie, Richard T. and Bollerslev, Tim and Mikkelsen, Hans Ole},
  title   = {Fractionally Integrated Generalized Autoregressive Conditional Heteroskedasticity},
  journal = {Journal of Econometrics},
  volume  = {74},
  number  = {1},
  pages   = {3--30},
  year    = {1996},
}

@article{barndorff2002econometric,
  author  = {Barndorff-Nielsen, Ole E. and Shephard, Neil},
  title   = {Econometric Analysis of Realized Volatility and Its Use in Estimating Stochastic Volatility Models},
  journal = {Journal of the Royal Statistical Society: Series B},
  volume  = {64},
  number  = {2},
  pages   = {253--280},
  year    = {2002},
}

@article{bennedsen2021decoupling,
  author  = {Bennedsen, Mikkel and Lunde, Asger and Pakkanen, Mikko S.},
  title   = {Decoupling the Short- and Long-Term Behavior of Stochastic Volatility},
  journal = {Journal of Financial Econometrics},
  volume  = {20},
  number  = {5},
  pages   = {961--1006},
  year    = {2021},
}

@article{bollerslev1986generalized,
  author  = {Bollerslev, Tim},
  title   = {Generalized Autoregressive Conditional Heteroskedasticity},
  journal = {Journal of Econometrics},
  volume  = {31},
  number  = {3},
  pages   = {307--327},
  year    = {1986},
}

@article{breiman2001random,
  author  = {Breiman, Leo},
  title   = {Random Forests},
  journal = {Machine Learning},
  volume  = {45},
  number  = {1},
  pages   = {5--32},
  year    = {2001},
}

@article{corsi2009simple,
  author  = {Corsi, Fulvio},
  title   = {A Simple Approximate Long-Memory Model of Realized Volatility},
  journal = {Journal of Financial Econometrics},
  volume  = {7},
  number  = {2},
  pages   = {174--196},
  year    = {2009},
}

@article{diebold1995comparing,
  author  = {Diebold, Francis X. and Mariano, Roberto S.},
  title   = {Comparing Predictive Accuracy},
  journal = {Journal of Business and Economic Statistics},
  volume  = {13},
  number  = {3},
  pages   = {253--263},
  year    = {1995},
}

@article{ding1993long,
  author  = {Ding, Zhuanxin and Granger, Clive W. J. and Engle, Robert F.},
  title   = {A Long Memory Property of Stock Market Returns and a New Model},
  journal = {Journal of Empirical Finance},
  volume  = {1},
  number  = {1},
  pages   = {83--106},
  year    = {1993},
}

@article{engle1982autoregressive,
  author  = {Engle, Robert F.},
  title   = {Autoregressive Conditional Heteroscedasticity with Estimates of the Variance of United Kingdom Inflation},
  journal = {Econometrica},
  volume  = {50},
  number  = {4},
  pages   = {987--1007},
  year    = {1982},
}

@article{friedman2001greedy,
  author  = {Friedman, Jerome H.},
  title   = {Greedy Function Approximation: A Gradient Boosting Machine},
  journal = {Annals of Statistics},
  volume  = {29},
  number  = {5},
  pages   = {1189--1232},
  year    = {2001},
}

@article{gatheral2018volatility,
  author  = {Gatheral, Jim and Jaisson, Thibault and Rosenbaum, Mathieu},
  title   = {Volatility Is Rough},
  journal = {Quantitative Finance},
  volume  = {18},
  number  = {6},
  pages   = {933--949},
  year    = {2018},
}

@article{geweke1983estimation,
  author  = {Geweke, John and Porter-Hudak, Susan},
  title   = {The Estimation and Application of Long Memory Time Series Models},
  journal = {Journal of Time Series Analysis},
  volume  = {4},
  number  = {4},
  pages   = {221--238},
  year    = {1983},
}

@article{granger1980introduction,
  author  = {Granger, Clive W. J. and Joyeux, Roselyne},
  title   = {An Introduction to Long-Memory Time Series Models and Fractional Differencing},
  journal = {Journal of Time Series Analysis},
  volume  = {1},
  number  = {1},
  pages   = {15--29},
  year    = {1980},
}

@article{gu2020empirical,
  author  = {Gu, Shihao and Kelly, Bryan and Xiu, Dacheng},
  title   = {Empirical Asset Pricing via Machine Learning},
  journal = {Review of Financial Studies},
  volume  = {33},
  number  = {5},
  pages   = {2223--2273},
  year    = {2020},
}

@book{hastie2009elements,
  author    = {Hastie, Trevor and Tibshirani, Robert and Friedman, Jerome},
  title     = {The Elements of Statistical Learning: Data Mining, Inference, and Prediction},
  edition   = {2},
  publisher = {Springer},
  year      = {2009},
}

@article{hosking1981fractional,
  author  = {Hosking, J. R. M.},
  title   = {Fractional Differencing},
  journal = {Biometrika},
  volume  = {68},
  number  = {1},
  pages   = {165--176},
  year    = {1981},
}

@article{mikosch2004nonstationarities,
  author  = {Mikosch, Thomas and St\u{a}ric\u{a}, C\u{a}t\u{a}lin},
  title   = {Nonstationarities in Financial Time Series, the Long-Range Dependence, and the {IGARCH} Effects},
  journal = {Review of Economics and Statistics},
  volume  = {86},
  number  = {1},
  pages   = {378--390},
  year    = {2004},
}

@article{muller1997volatilities,
  author  = {M\"uller, Ulrich A. and Dacorogna, Michel M. and Dav\'e, Rakhal D. and Olsen, Richard B. and Pictet, Olivier V. and von Weizs\"acker, Jakob E.},
  title   = {Volatilities of Different Time Resolutions---Analyzing the Dynamics of Market Components},
  journal = {Journal of Empirical Finance},
  volume  = {4},
  number  = {2--3},
  pages   = {213--239},
  year    = {1997},
}

@article{robinson1995gaussian,
  author  = {Robinson, Peter M.},
  title   = {Gaussian Semiparametric Estimation of Long Range Dependence},
  journal = {Annals of Statistics},
  volume  = {23},
  number  = {5},
  pages   = {1630--1661},
  year    = {1995},
}

@article{harvey1997testing,
  author  = {Harvey, David and Leybourne, Stephen and Newbold, Paul},
  title   = {Testing the Equality of Prediction Mean Squared Errors},
  journal = {International Journal of Forecasting},
  volume  = {13},
  number  = {2},
  pages   = {281--291},
  year    = {1997},
}

@article{moreira2017volatility,
  author  = {Moreira, Alan and Muir, Tyler},
  title   = {Volatility-Managed Portfolios},
  journal = {Journal of Finance},
  volume  = {72},
  number  = {4},
  pages   = {1611--1644},
  year    = {2017},
}

@article{patton2011volatility,
  author  = {Patton, Andrew J.},
  title   = {Volatility Forecast Comparison Using Imperfect Volatility Proxies},
  journal = {Journal of Econometrics},
  year    = {2011},
  volume  = {160},
  number  = {1},
  pages   = {246--256},
  doi     = {10.1016/j.jeconom.2010.03.034}
}

\end{document}